\documentclass[aps, pre, twocolumn, notitlepage, a4paper, floatfix, nofootinbib, longbibliography]{revtex4-1}

\usepackage[utf8]{inputenc}
\usepackage[english]{babel}
\usepackage{graphicx}
\usepackage{amssymb}
\usepackage{amsmath}
\usepackage{enumitem}
\usepackage{bm}
\usepackage{hyperref}
\usepackage{float}
\usepackage{amsmath}
\usepackage[percent]{overpic}
\usepackage{color}
\usepackage{overpic}
\usepackage{grffile}

\usepackage[normalem]{ulem}

\graphicspath{{./figures/}}

\newcommand{\avg}[1]{{\left<#1\right>}}

\newcommand{\A}{\bm{A}}
\newcommand{\bb}{\bm{b}}
\newcommand{\p}{{\bm{p}}}
\newcommand{\w}{{\bm{\omega}}}
\newcommand{\m}{{\bm{m}}}
\newcommand{\ee}{{\mathrm{e}}}

\begin{document}

\title{Null models for multi-optimized large-scale network structures}
\author{Sebastian Morel-Balbi}
\email{smb81@bath.ac.uk}
\affiliation{Department of Mathematical Sciences, University of Bath, Claverton Down, Bath BA2 7AY, United Kingdom}
\author{Tiago P. Peixoto}
\email{peixotot@ceu.edu}
\affiliation{Department of Network and Data Science, Central European University, H-1051 Budapest, Hungary}
\affiliation{ISI Foundation, Via Chisola 5, 10126 Torino, Italy}
\affiliation{Department of Mathematical Sciences, University of Bath, Claverton Down, Bath BA2 7AY, United Kingdom}

\begin{abstract}
  We study the emerging large-scale structures in networks subject to
  selective pressures that simultaneously drive towards higher
  modularity and robustness against random failures. We construct
  maximum-entropy null models that isolate the effects of the joint
  optimization on the network structure from any kind of evolutionary
  dynamics. Our analysis reveals a rich phase diagram of optimized
  structures, composed of many combinations of modular, core-periphery
  and bipartite patterns. Furthermore, we observe parameter regions
  where the simultaneous optimization can be either synergistic or
  antagonistic, with the improvement of one criterion directly aiding or
  hindering the other, respectively. Our results show how interactions
  between different selective pressures can be pivotal in determining
  the emerging network structure, and that these interactions can be
  captured by simple network models.
\end{abstract}

\maketitle

\section{Introduction}

The observed large-scale structure of network systems emerges as the
outcome of various kinds of generative processes, which tend to vary
substantially depending on their empirical context. Nevertheless, in a
large class of network formation mechanisms, in particular in
biological, engineering and technological settings, an important driving
force is the fitness to a specific
purpose~\cite{sole_model_2002,cancho_optimization_2003,paul_optimization_2004,gastner_optimal_2006,
barthelemy_optimal_2006, wagner_road_2007},
e.g. the survival of an individual, the efficiency of a production line,
or the capacity of a transportation system. This results in a selective
pressure towards particular network structures, depending on the kind of
fitness that is desired. However, in realistic scenarios, selective
pressures occur in combination with other kinds of dynamical rules,
exogenous constraints and historical artefacts. Furthermore, a given
system may be subject to multiple selective pressures at the same time,
e.g. it may need to run efficiently while being simultaneously robust to
errors or damage. Since very seldom do we get to observe any given
process of formation in detail, we are forced to disentangle these
different driving forces from each other based only on the structural
patterns they produce.

In this work, we contribute to the disentangling effort by constructing
\emph{null models} of optimized
networks~\cite{peixoto_evolution_2012}. These models correspond to
network ensembles that possess a pre-specified level of fitness, but
otherwise are maximally random. By investigating the emerging structural
features in these models, we are able to understand the inherent effect
a particular kind of fitness criterion has on the structure of the
network, without the interference of any other kind of constraint. We
can also combine multiple fitness criteria together to determine how
they interact with each other in determining the preferred network
structure. This gives us a controlled platform to delineate the effects
of different kinds of selective pressures on network structure in a
principled manner.

In the following we will employ this approach to investigate two central
properties of networked systems, namely the robustness of a network
against the random failure of its
components~\cite{callaway_network_2000}, and its
modularity~\cite{girvan_community_2002}, characterized by the existence
of groups of nodes that are more connected among themselves than with
the rest of the network. Robustness to failure is believed to play a key
role in infrastructure~\cite{buldyrev_catastrophic_2010} as well as
technological networks such as the
internet~\cite{cohen_resilience_2000}, but also on biological
systems~\cite{deutscher_multiple_2006}. Modularity, on the other hand,
has been associated with the adaptability of biological
networks~\cite{wagner_road_2007}, and is a necessary ingredient for the
scheduling of interdependent processes with minimal amount of
communication~\cite{buluc_recent_2016}. By enforcing these two
optimization criteria simultaneously, we analyse which large-scale
network structures are most likely to emerge as a result of their
interaction. Our main result is the identification of a series of phase
transitions at which the optimal structure of the network changes in
response to the varying selective pressures. We also identify regions in
the parameter space where the interplay between the selective pressures
gives rise to synergistic effects, i.e. one kind of fitness pressure
contributes to the second, such that it becomes easier to optimize for
both at once, as well as antagonistic effects, where both optimizations
compete against each other.

The work is divided as follows. We begin in Sec.~\ref{sec:model} by
introducing our modelling framework.  In Sec.~\ref{sec:criteria}, we
apply our framework to network ensembles subject to varying degrees of
selective pressures in favor of robustness against random failures as
well as modularity. We begin by considering each fitness criteria
separately and subsequently combine them to analyse the effects that
their interaction has on the emerging network structures. Finally, in
Sec.~\ref{sec:conclusion}, we draw our conclusions.

\section{Null models of optimized modular networks}
\label{sec:model}

We approach the problem of characterizing network structures via
generative models. This means that instead of describing individual
networks, we are interested in formulating network ensembles, such that
the probability of observing a given network is associated with its
particular fitness value, given a predefined fitness criterion. There
are many ways to address this problem, but here we constrain ourselves
to networks that exhibit modular structure, i.e. the nodes are divided
into groups, which share a similar role in the network structure. More
specifically, we consider networks that are generated from the
stochastic block model (SBM)~\cite{holland_stochastic_1983,
wasserman_stochastic_1987, karrer_Stochastic_2011}, where $N$ nodes are
divided into $B$ groups, such that to each node $i$ is given a group
membership label $b_i\in \{1,\ldots, B\}$, and an edge between a node in
group $r$ and another in group $s$ exists with probability
$p_{rs}$. This yields a network ensemble where a network $\bm A$ occurs
with probability
\begin{equation}\label{eq:sbm}
  P(\A | \bb, \p) = \prod_{i<j}p_{b_i,b_j}^{A_{ij}}(1-p_{b_i,b_j})^{1-A_{ij}},
\end{equation}
where $A_{ij}=1$ if an edge exists between nodes $(i,j)$, or $A_{ij}=0$
otherwise. Although this is just one of a large set of possible network
ensembles, the SBM is capable of capturing arbitrary mixing patterns
between groups by appropriate choices of the matrix $\p$, and if the
number of groups is increased it can account for arbitrarily elaborate
network structures~\cite{olhede_network_2014}. In fact, setting $B=N$
means that the probability of each edge can be individually controlled,
although we will constrain ourselves to the situation where $B\ll N$ and
the network is composed of a relatively small number of
modules. Although this does not give us the full breadth of all possible
network structures --- in particular we lack the ability of describing
the details of the network structure at a local level, e.g. by
stipulating desired propensities of observing triangles or other small
subgraphs --- as we will see, this is a sufficiently flexible framework
to express the kind of null models we have in mind.

For a given arbitrary fitness function $R(\A)$, which maps a network to
a scalar fitness value, the average fitness over the SBM ensemble is
then given by
\begin{equation}
  R(\bb, \p) = \sum_{\A}R(\A)P(\A | \bb, \p).
\end{equation}
Based on such a function, we could in principle proceed by finding the
SBM parameters $\bb$ and $\p$ such that the mean fitness $R(\bb,\p)$ is
maximized, and in this way finding how a fitness criterion favors
certain patterns of network structures. However, this kind of
optimization problem is ill-defined in the general case, as many
parameter choices yield the same optimal fitness value. Therefore, we
formulate our question differently. Instead of optimizing the mean
fitness $R(\bb,\p)$, we impose its value as a pre-determined parameter,
and we select the SBM parameters that yield the most random network
ensemble, and therefore is the most agnostic about the unimportant
properties of the network structure. More formally, this means we employ
the principle of maximum entropy~\cite{jaynes_probability_2003}, such
that for any imposed fitness value $R(\bb,\p)=R^*$, the choice of the
model parameters $\bb$ and $\p$ from all those that fulfill this
constraint is the one that maximizes the ensemble
entropy~\cite{bianconi_entropy_2009},
\begin{equation}\label{eq:entropy}
  \Sigma(\bb, \p) = -\sum_{\A}P(\A | \bb, \p)\ln P(\A | \bb, \p).
\end{equation}
In this way, if we specify a set of fitness functions $\{R_i(\bb, \p)\}$ and
their imposed set of values $\{R_i^*\}$, we are interested in the following
constrained optimization problem
\begin{equation}\label{eq:opt}
  \hat\bb, \hat\p = \underset{\bb,\p}{\operatorname{argmax}}\;\Sigma(\bb, \p),\; \text{ subject to }
  R_i(\bb,\p) = R^*_i\;\; \forall i.
\end{equation}
The SBM parameters obtained in this way can be interpreted as \emph{null
models} of networks, which contain only the most essential ingredients
to achieve the pre-specified values of fitness, and otherwise are
maximally random. The imposed fitness values themselves can be increased
arbitrarily, to achieve any level of optimized structures, as we will
show.

We can compute the entropy of the SBM ensemble by substituting
Eq.~\ref{eq:sbm} into Eq.~\ref{eq:entropy}, which yields~\cite{peixoto_entropy_2012}
\begin{equation}
  \Sigma(\bb, \p) =\sum_{r<s}n_rn_sH_{\text{b}}(p_{rs}) + \sum_{r}\frac{n_r(n_r-1)}{2}H_{\text{b}}(p_{rr}),
\end{equation}
where $n_r=\sum_i\delta_{b_i,r}$ is the number of nodes in group $r$,
and $H_{\text{b}}(x)=-x\ln x -(1-x)\ln (1-x)$ is the binary entropy
function. This can be further simplified if we take into account that
most networks in the real world are sparse with $p_{rs}=O(1/N)$, so that
using $H_{\text{b}}(x)=-x\ln x + x + O(x^2)$, and taking the limit $N\gg
1$ we obtain
\begin{equation}
  \Sigma(\bb, \p) = -\frac{1}{2}\sum_{rs}n_rn_s\left(p_{rs}\ln p_{rs} - p_{rs}\right).
\end{equation}
For some choices of fitness functions, arbitrarily high fitness values
can be obtained simply by increasing the network density. In order to
differentiate between the effect of increased density and favored mixing
patterns, we will take the average degree $\avg{k} =
\sum_{rs}n_rn_sp_{rs}/N$ as an external parameter not subject to
optimization. With this in mind, it will be useful for our calculations
to use the following re-parametrization over intensive variables,
\begin{equation}
  \omega_r = \frac{n_r}{N}, \qquad m_{rs} = \frac{n_rn_sp_{rs}}{N\avg{k}}.
\end{equation}
Note that the above implies the normalization $\sum_r\omega_r=1$
and $\sum_{rs}m_{rs}=1$. Given this choice, the ensemble entropy can be
written as
\begin{equation}
\Sigma(\w, \m) = -\frac{\avg{k}N}{2}\sum_{rs}m_{rs}\ln \frac{m_{rs}}{\omega_r\omega_s} + \frac{\avg{k}{N}}{2}
\end{equation}
Note that we no longer reference the actual partition $\bb$ itself, but
rather the fraction of nodes $\omega_r$ that belong to a given group
$r$, since these are the relevant macroscopic quantities as $N\gg 1$.

Based on the above model parametrization, we can perform the constrained
optimization of Eq.~\ref{eq:opt} by employing the method of Lagrange
multipliers, which involves finding the saddle points of the Lagrangian
function
\begin{equation}
  \Lambda(\w, \m, \bm{\beta}) = \Sigma (\w, \m) + \sum_i \beta_i \left[R_i(\w, \m) - R_i^*\right],
\end{equation}
where $\beta_i$ are the Lagrange multipliers that enforce each
constraint. This means we need to find $\w$, $\m$, and $\bm{\beta}$ such
that the gradient of $\Lambda$ is zero,
i.e. $\partial \Lambda(\w, \m, \bm{\beta})/\partial \omega_r = \partial
\Lambda(\w, \m, \bm{\beta})/\partial m_{rs} = \partial \Lambda(\w, \m,
\bm{\beta})/\partial \beta_i = 0$. Note that the last derivative yields
simply the equation $R_i(\w, \m) = R_i^*$, which means that the problem
of fixing $R_i^*$ and finding $\w, \m, \bm{\beta}$ is equivalent to
first taking $\bm\beta$ as fixed parameters and minimizing the function
\begin{equation}\label{eq:fe}
  \mathcal{F}(\w, \m) = -\sum_i \beta_i R_i(\w, \m) - \Sigma (\w, \m),
\end{equation}
with respect to $\w$ and $\m$ alone and then varying $\bm\beta$ until we
obtain $R_i(\w, \m) = R_i^*$.

The above formulation puts us in a standard setting in equilibrium
statistical physics, as the function $\mathcal{F}(\w, \m)$ can be
interpreted as the \emph{free energy} of the network ensemble where the
sum $-\sum_i\beta_i R_i(\w, \m)$ plays the role of the mean
energy. Following this analogy, the values of $\beta_i$ play the role of
inverse temperatures, or perhaps more appropriately to our setting,
\emph{selective pressures}, which if increased cause the corresponding
energy functions to decrease (and thus the fitness values to increase),
and thus settling on a particular balance between energy and entropy.

To summarize, our protocol to generate null network models is as
follows:
\begin{enumerate}
\item We establish a set of fitness functions $\{R_i(\w,\m)\}$.
\item Given a choice of selective pressures $\{\beta_i\}$ we find the
parameters $\w,\m$ which minimize the free energy $\mathcal{F}(\w, \m)$ of
Eq.~\ref{eq:fe}.
\item We vary the values $\{\beta_i\}$ to investigate the trade-off
      between competing fitness functions as well as entropy.
\end{enumerate}
The constrained optimization of step 2 is the most central part of our
approach. Although it is straightforward to compute the gradient of the
entropy $\Sigma(\w,\m)$ analytically, in the general case this will not
be possible for arbitrary fitness functions $R_i(\w,\m)$, and even when
it is, setting the gradient of $\mathcal{F}(\w,\m)$ to zero usually just
yields an implicit system of nonlinear equations that cannot be solved
in closed form. Therefore, in the following we will proceed by
performing the minimization numerically, via the L-BFGS-B conjugate
gradient descent algorithm~\cite{byrd_limited_1995}, using automatic
differentiation~\cite{baydin_automatic_2015} whenever the gradient
cannot be obtained in closed form. As a final implementation note, the
used algorithms require us to convert step 2 into an unbounded
optimization problem, which we do via a simple exchange of variables
given by
\begin{equation}
  \omega_r = \frac{\ee^{\mu_r}}{\sum_s \ee^{\mu_s}}, \qquad
  m_{rs} = \frac{\ee^{\nu_{rs}}}{\sum_{tu}\ee^{\nu_{tu}}},
\end{equation}
with $\mu_r \in [-\infty, \infty]$ and $\nu_{rs} \in [-\infty, \infty]$,
which keep both $\omega_r$ an $m_{rs}$ bounded in the range
$[0,1]$, and enforces normalization.

\section{Fitness criteria}
\label{sec:criteria}

We consider two kinds of fitness criteria, namely the robustness against
random failures, and modularity. We begin by considering the criteria in
isolation, and we follow by combining them simultaneously.

\subsection{Robustness against random failures}

We consider a situation where a random fraction $1-\phi$ of the edges
are removed from the network, and we measure the fraction $S$ of nodes
that remain connected afterwards, forming a giant connected
component~\cite{newman_random_2001}. Following
Ref.~\cite{peixoto_evolution_2012}, we can compute this quantity for the
SBM by first defining $u_r$ to be the probability that a node in group
$r$ does not belong to the giant component via one of its neighbors,
which can be obtained by solving the set of equations
\begin{equation}\label{eq:ur}
   u_r = 1- \phi+ \phi\sum_s \frac{m_{rs}}{m_r}f_1^s(u_s),
\end{equation}
where $m_r = \sum_sm_{rs}$, and $f_1^r(z)=f_0^{r'}(z)/f_0^{r'}(1)$ is
the generating function of the excess degree distribution of nodes
belonging to group $r$, defined in terms of the corresponding degree
distribution generating function $f_0^r(z)$ given by
\begin{equation}
  f_0^r(z) = \sum_k p_k^r z^k
\end{equation}
where $p_k^r$ is the fraction of nodes in group $r$ that have degree
$k$, which for the SBM is a Poisson distribution with mean
$\kappa_r = \avg{k}m_r/\omega_r$, which means we have
\begin{equation}
        f_0^r(z) = f_1^r(z) = f^r(z) = \mathrm{e}^{\kappa_r(z - 1)}.
\end{equation}
After solving Eq.~\ref{eq:ur}, which can be done simply by repeated
iteration from a starting point $u_r < 1$ until a desired convergence
criterion, we can finally obtain $S$ via
\begin{equation}
  S = 1 - \sum_r\omega_r f_0^r(u_r).
\end{equation}
For any given SBM, the behavior of $S$ as a function of the fraction
$\phi$ of edges that are not removed is that we have $S=0$ for $\phi \in
[0, \phi^*]$, where $\phi^*$ is a critical value, so that for $\phi >
\phi^*$ we have a positive fraction of connected nodes $S>0$ that
increases continuously~\cite{peixoto_evolution_2012}.

If we now consider the fitness function $R(\w,\m) = S(\w,\m)$, our
resulting free energy becomes
\begin{equation} \label{eq:f-Sopt}
  \mathcal{F}(\w,\m) = -\beta_S S(\w,\m) - \Sigma(\w,\m).
\end{equation}
By minimizing the above function we find null network models that are
robust against random failures, with the robustness increasing for
higher $\beta_S$ values.

\begin{figure}
  \begin{tabular}{c}
    \begin{overpic}[width=\columnwidth]{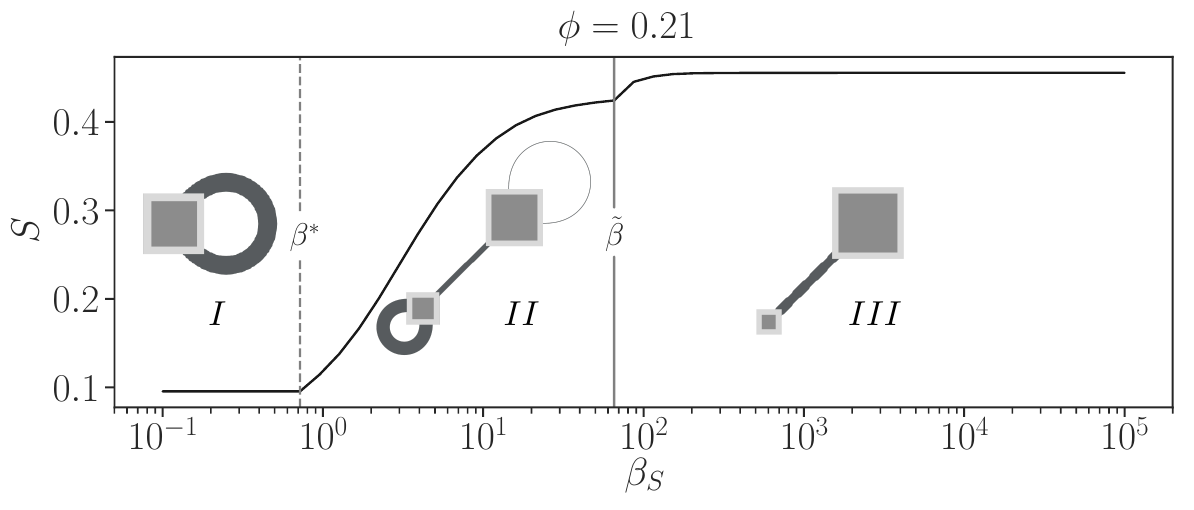}
      \put(0, 0){(a)}
    \end{overpic}\\
    \begin{overpic}[width=\columnwidth]{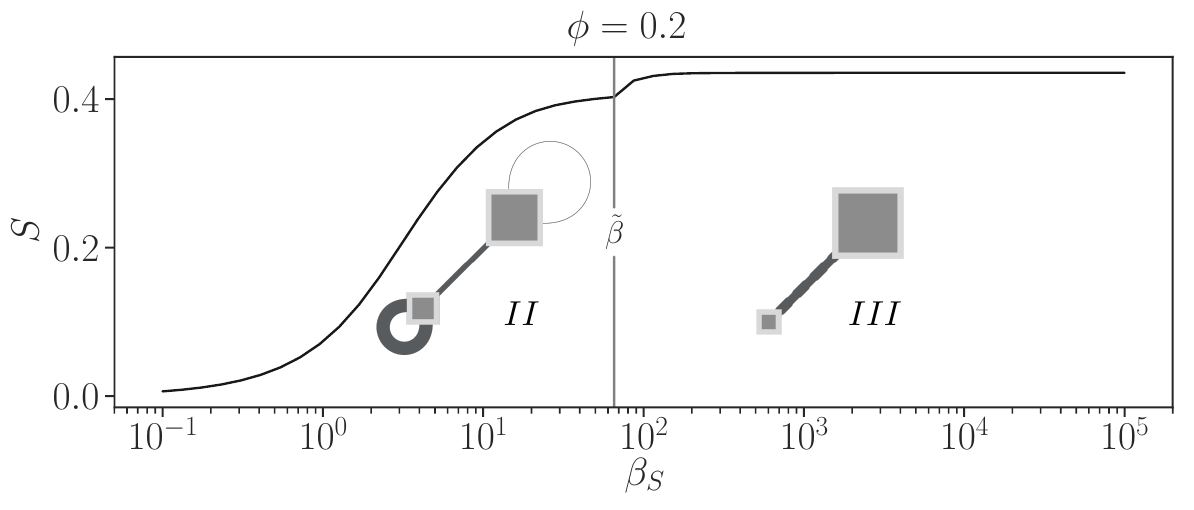}
      \put(0, 0){(b)}
    \end{overpic}\\
    \begin{overpic}[width=\columnwidth]{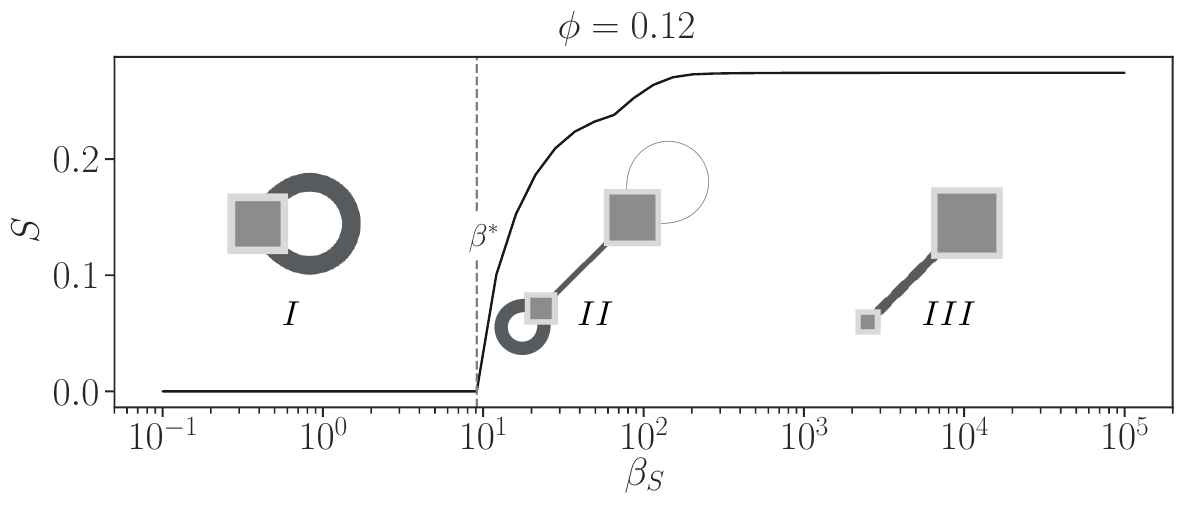}
      \put(0, 0){(c)}
    \end{overpic}
  \end{tabular}
  \caption{Relative size of the giant component $S$ as a function of the
    selective pressure for robustness to damage $\beta_S$ for different
    values of the edge dilution probability $\phi$.  The dashed vertical
    lines indicate the value $\beta_S = \beta^*$ at which we observe a
    transition from a random structure to a core-periphery one. The solid
    vertical lines indicate the value $\beta_S = \tilde{\beta}$ at which
    the network structure transitions from a core-periphery to a
    bipartite pattern. The optimized network structures are shown
    schematically in the insets, where each group corresponds to one of
    the groups of our model.\label{fig:S-vs-bS}}
\end{figure}

\begin{figure}
  \begin{tabular}{ccc}
  \begin{overpic}[width=.33\columnwidth]{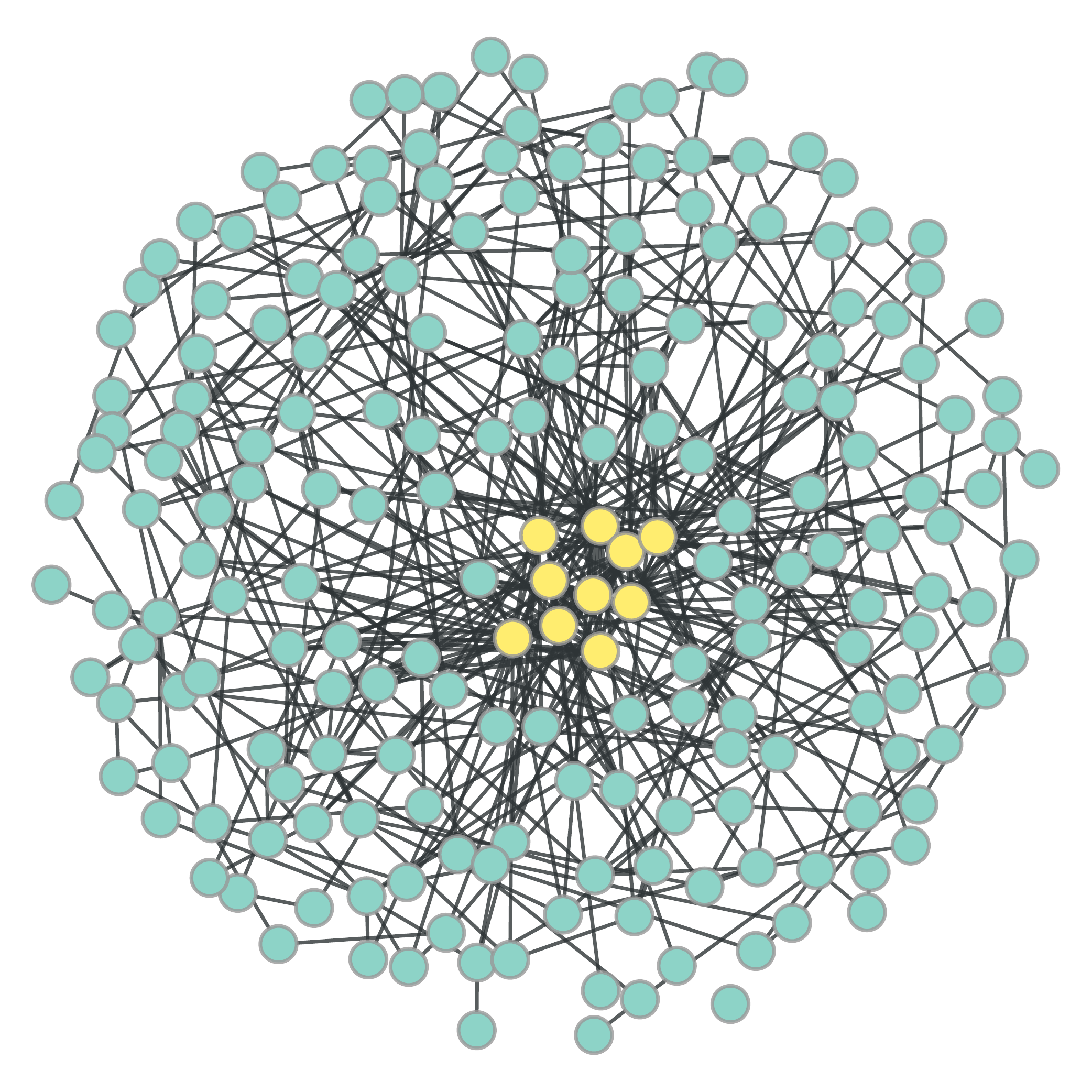}
  \end{overpic}&
  \begin{overpic}[width=.33\columnwidth]{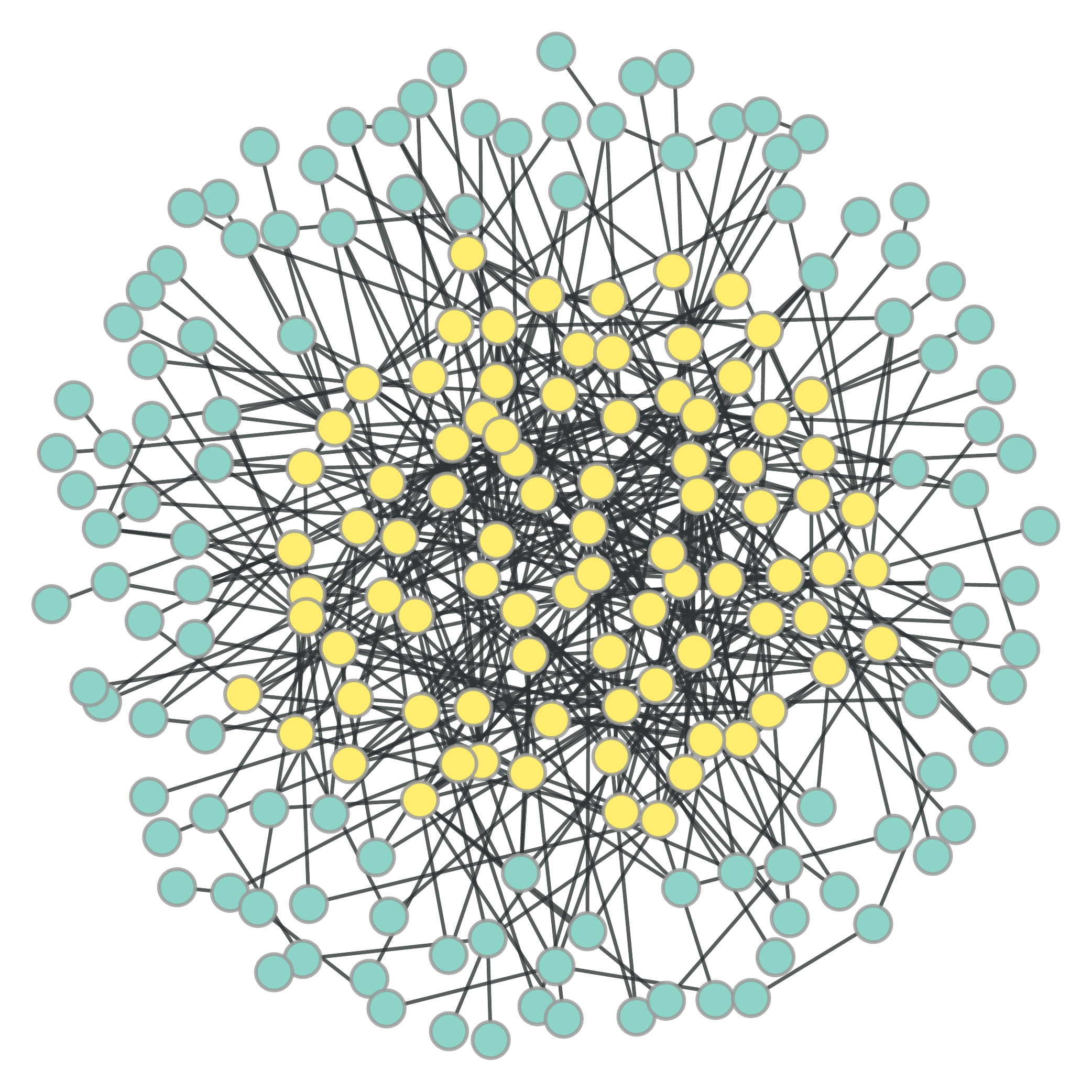}
  \end{overpic}&
  \begin{overpic}[width=.33\columnwidth]{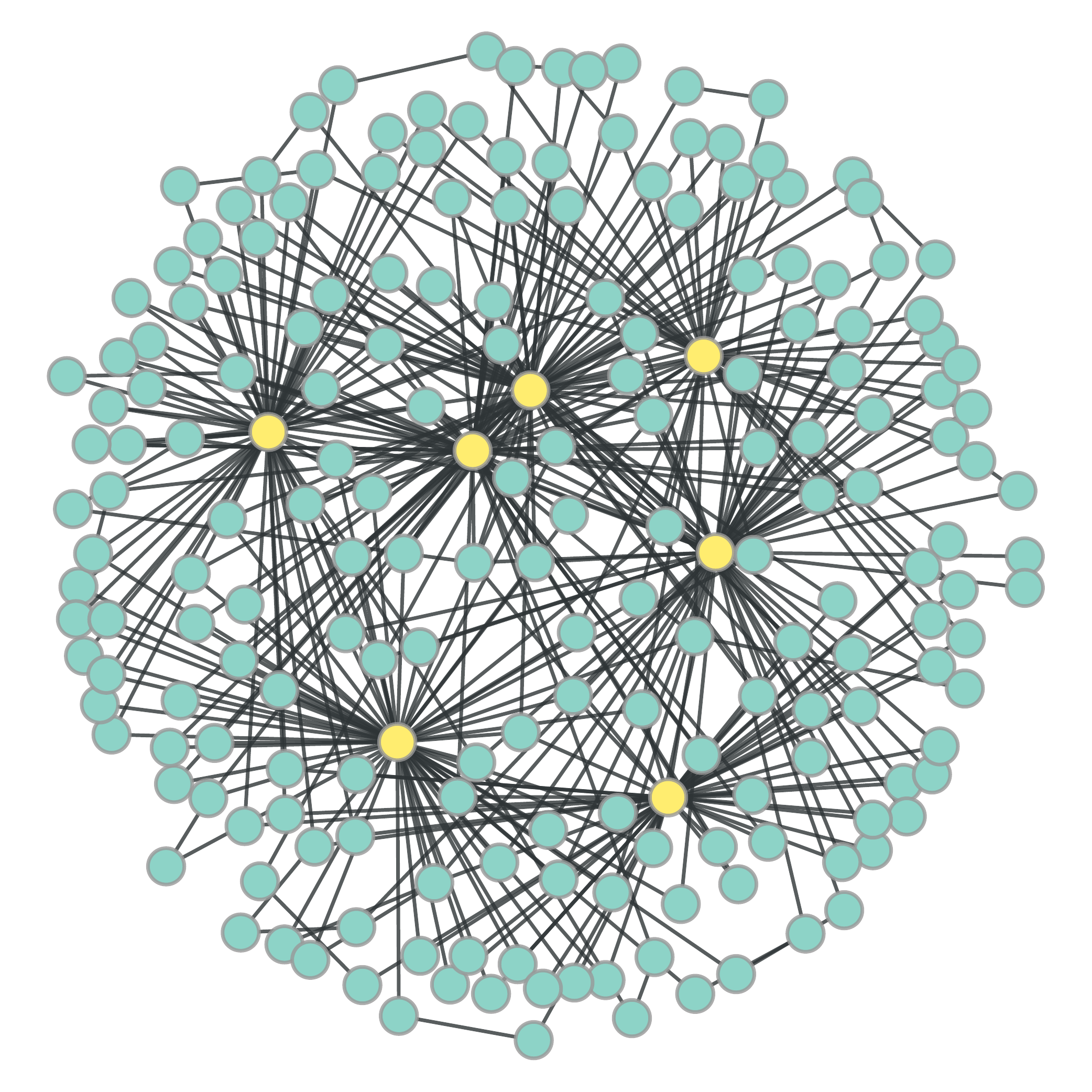}
  \end{overpic} \\
  (a) & (b) & (c)\\
  \multicolumn{3}{c}{{\LARGE $\xrightarrow[\beta_S]{\hspace*{6cm}}$}}
  \end{tabular}
 \caption{Ensemble samples depicting the typical evolution of the core-periphery
   structure as a function of the selective pressure $\beta_S$. (a) When
   the core-periphery structure first appears, it is composed of a small
   high-degree core. (b) As $\beta_S$ increases, the size of the core
   group becomes larger, (c) before eventually transitioning to a
   bipartite structure.
   \label{fig:S/low-phi-nw-samples}}
\end{figure}

\begin{figure}
  \begin{tabular}{cc}
  \begin{overpic}[width=.5\columnwidth]{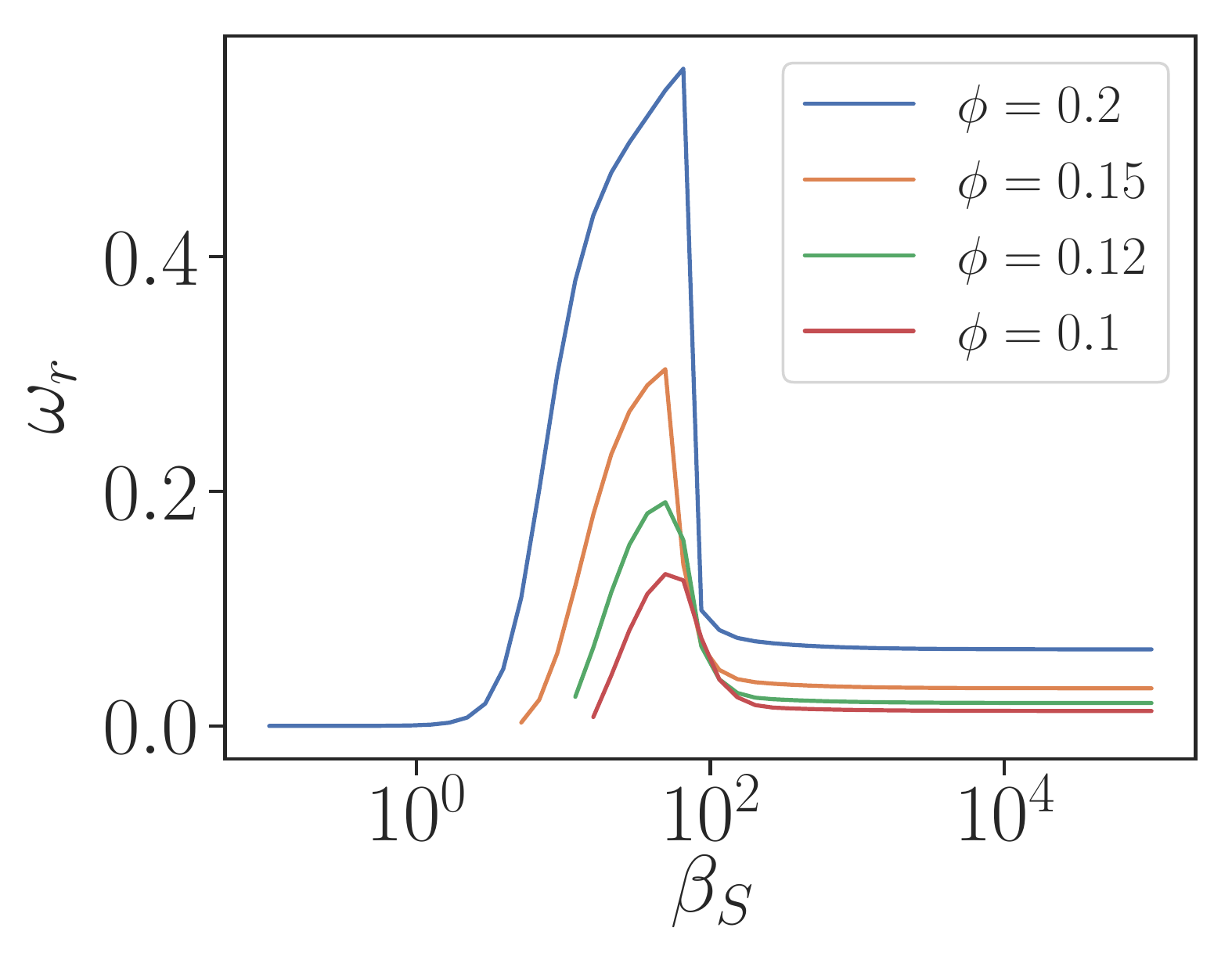}
    \put(0,0){(a)}
  \end{overpic}&
  \begin{overpic}[width=.5\columnwidth]{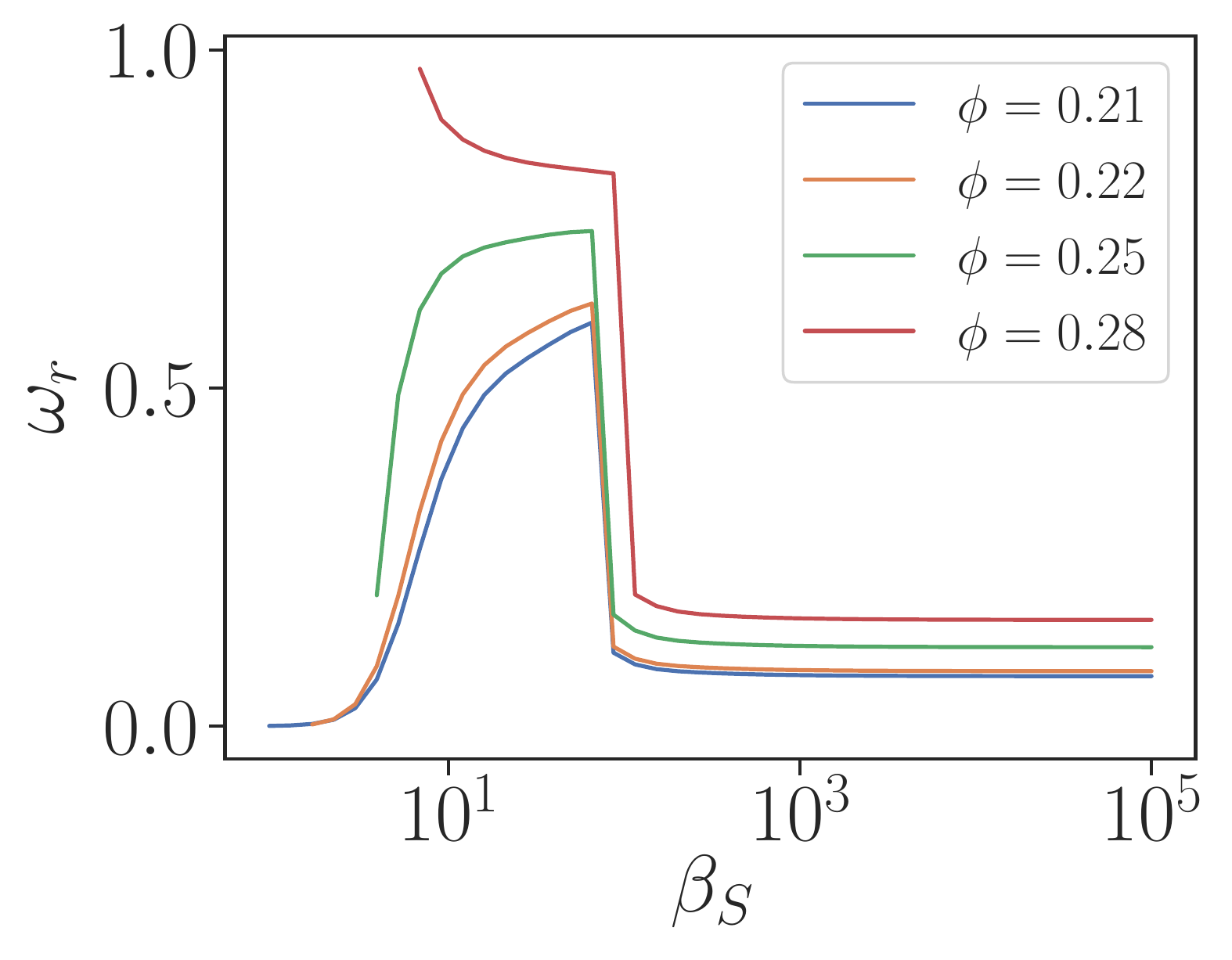}
    \put(0,0){(b)}
  \end{overpic}\\
  \begin{overpic}[width=.5\columnwidth]{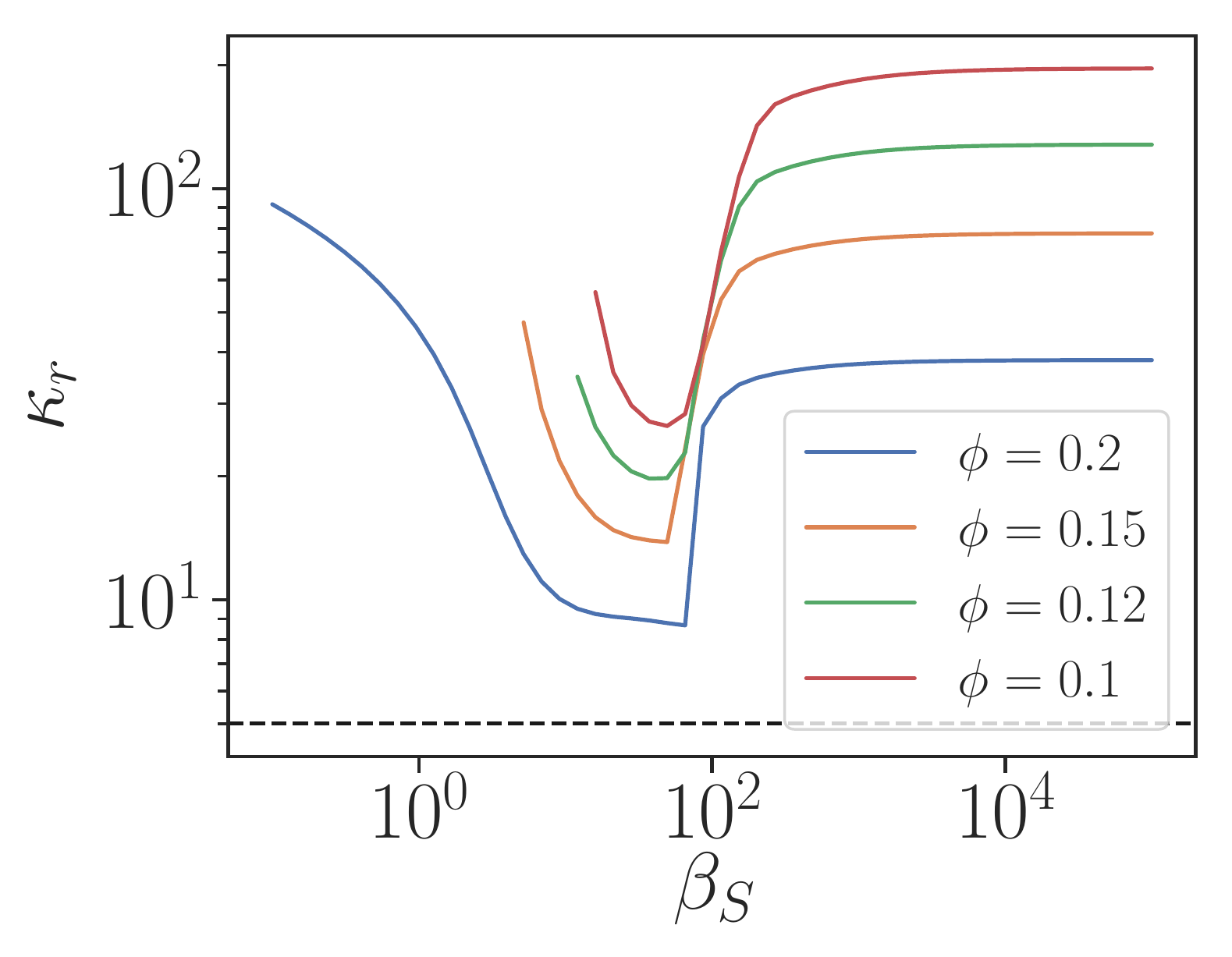}
    \put(0,0){(c)}
  \end{overpic}&
  \begin{overpic}[width=.5\columnwidth]{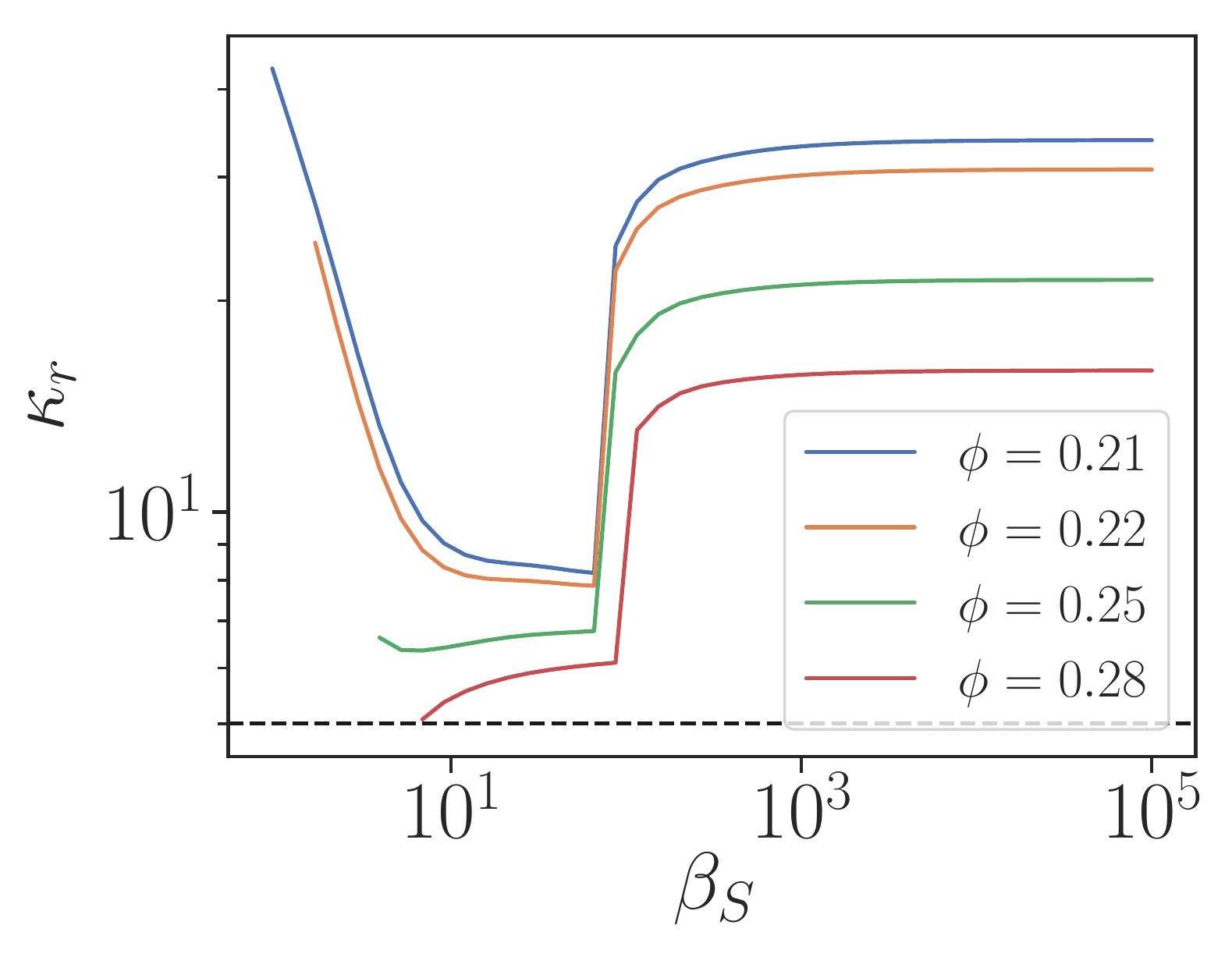}
    \put(0,0){(d)}
  \end{overpic}
  \end{tabular}
  \caption{
    Fraction of nodes and average degree of the core groups as a
    function of the selective pressure $\beta_S$. Panels on the left
    display curves for values of $\phi \leq \phi_c$. Panels on the right
    display curves for values of $\phi > \phi_c$. The black dashed line
    in the plots for $\kappa_r$ indicates the average degree of the
    network, which has been externally fixed to $\avg{k} =
    5$.\label{fig:S/core-sizes}}
\end{figure}

\begin{figure}
  \begin{tabular}{cc}
  \begin{overpic}[width=.5\columnwidth]{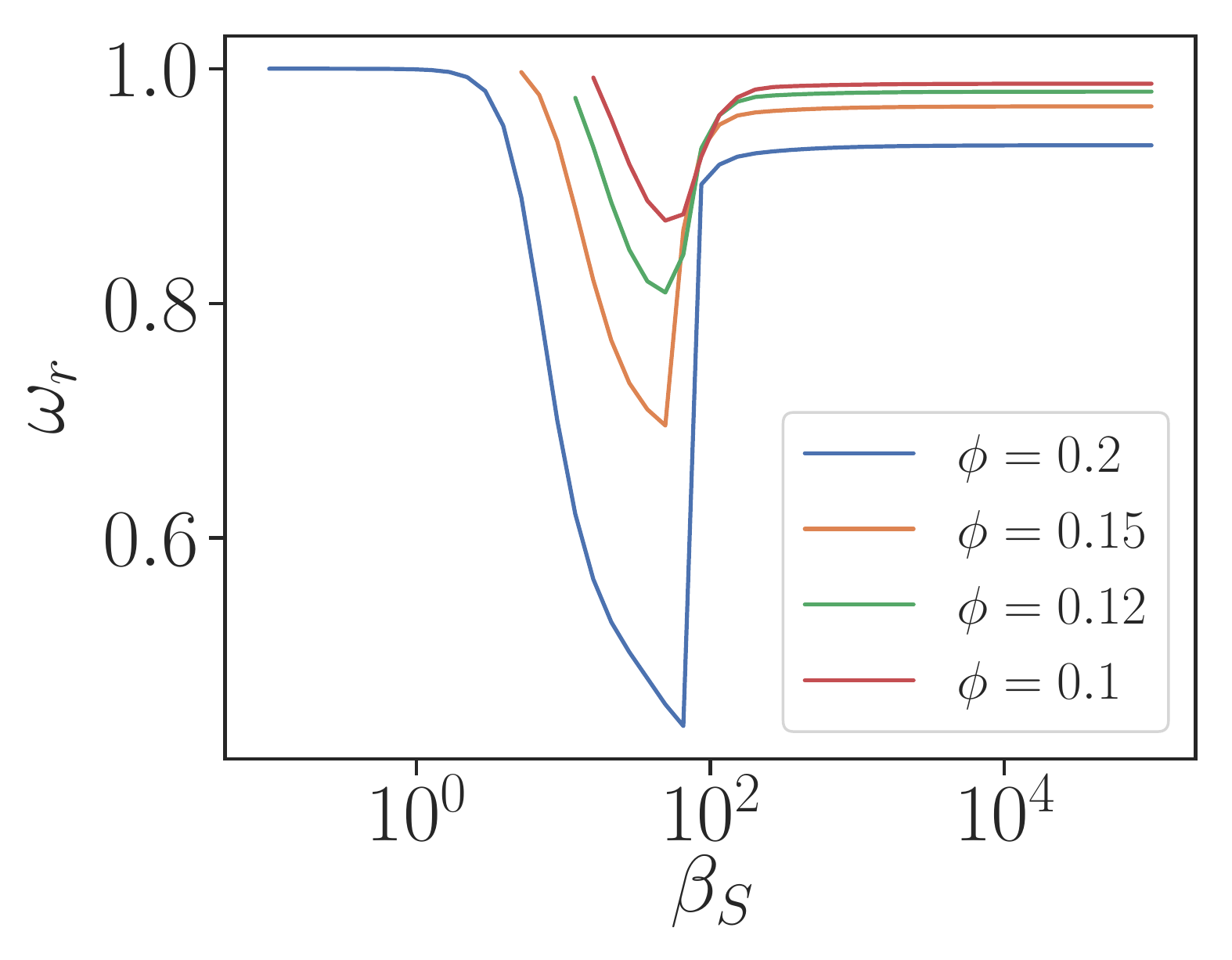}
    \put(0,0){(a)}
  \end{overpic}&
  \begin{overpic}[width=.5\columnwidth]{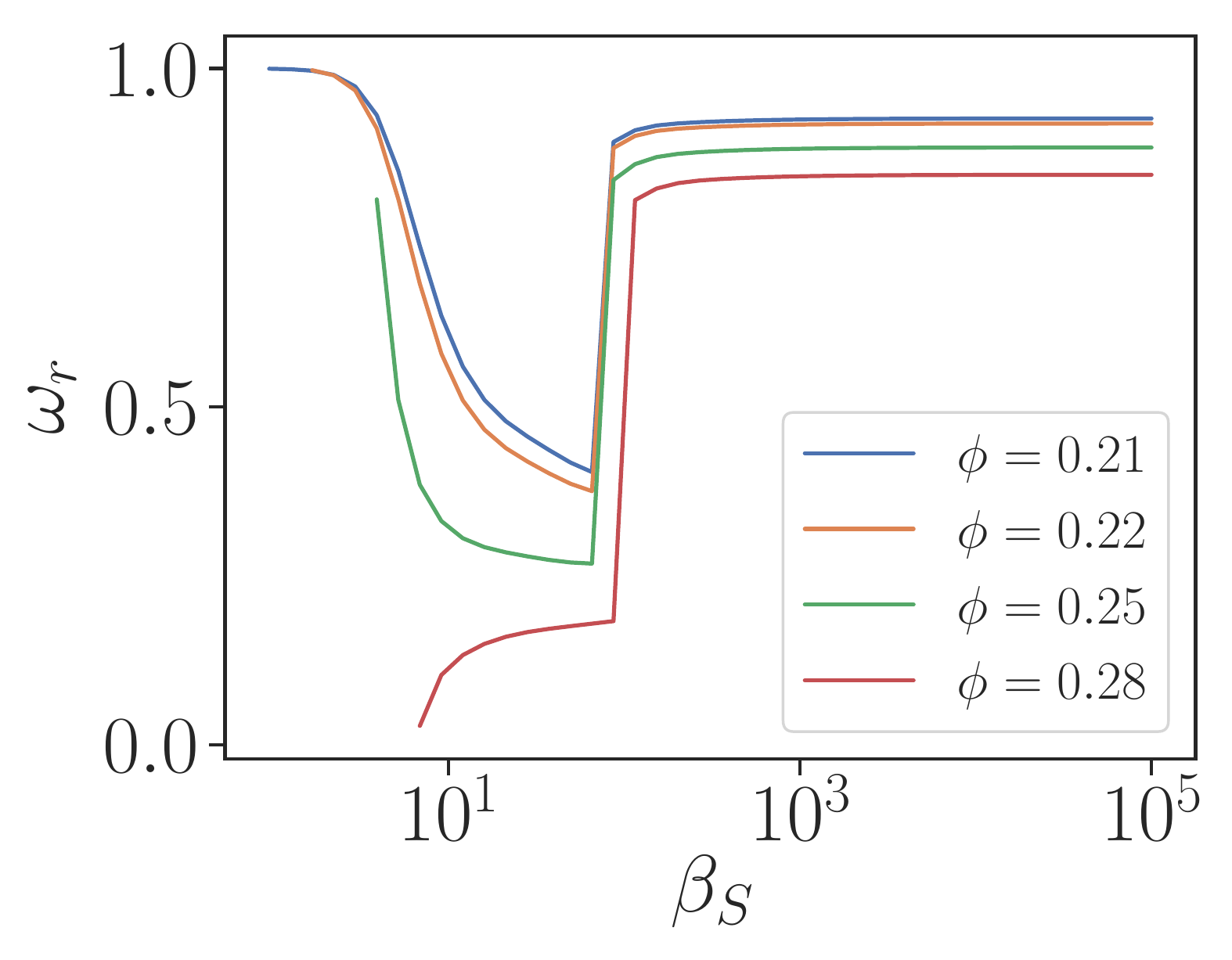}
    \put(0,0){(b)}
  \end{overpic}\\
  \begin{overpic}[width=.5\columnwidth]{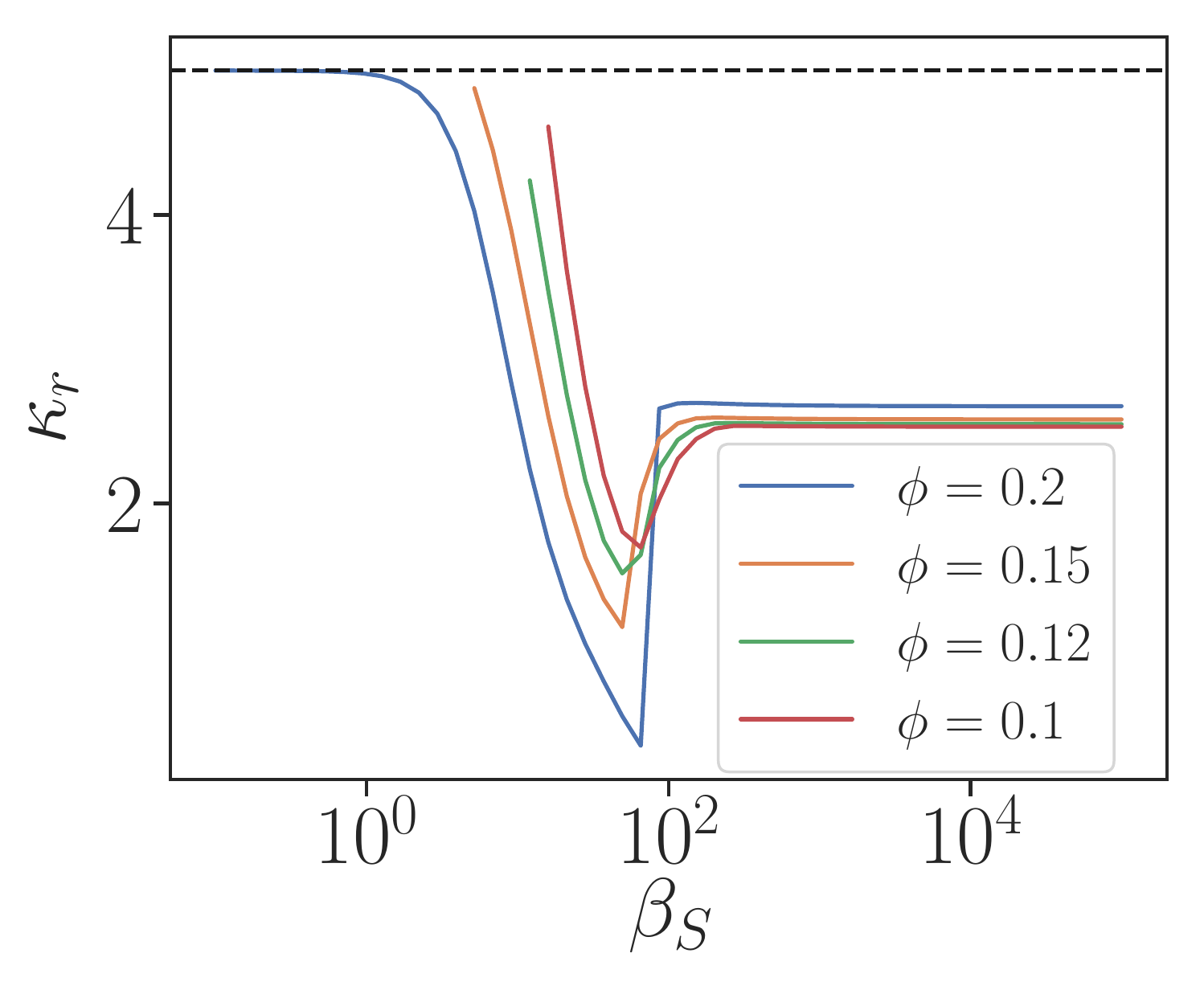}
    \put(0,0){(c)}
  \end{overpic}&
  \begin{overpic}[width=.5\columnwidth]{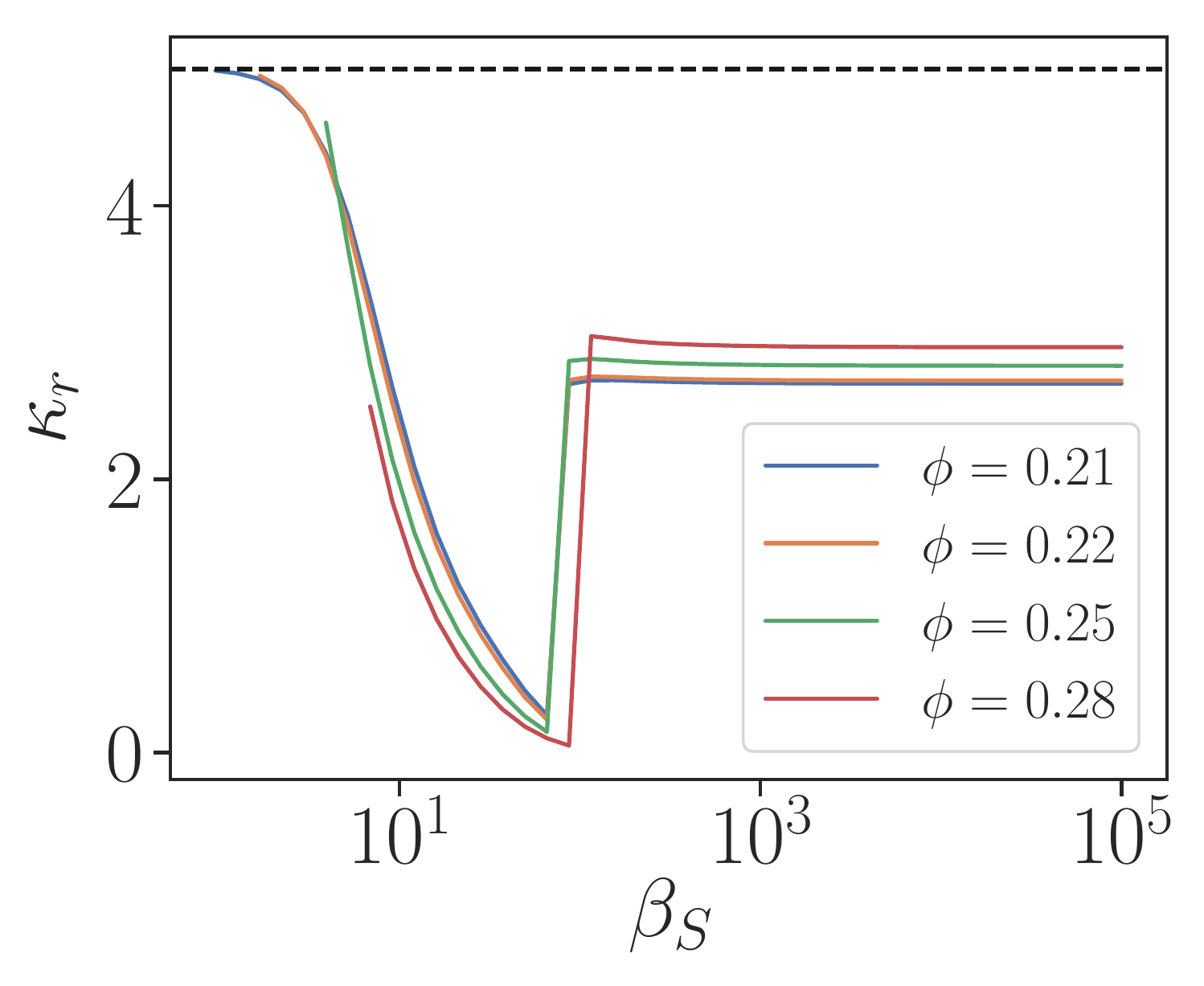}
    \put(0,0){(d)}
  \end{overpic}
  \end{tabular}
  \caption{
    Fraction of nodes and average degree of the periphery groups as a
    function of the selective pressure $\beta_S$. Panels on the left
    display curves for values of $\phi \leq \phi_c$. Panels on the right
    display curves for values of $\phi > \phi_c$. The black dashed line
    in the plots for $\kappa_r$ indicates the average degree of the
    network, which has been externally fixed to $\avg{k} =
    5$.\label{fig:S/periphery-sizes}}
\end{figure}

\begin{figure}
\begin{tabular}{c}
  \begin{overpic}[width=\columnwidth]{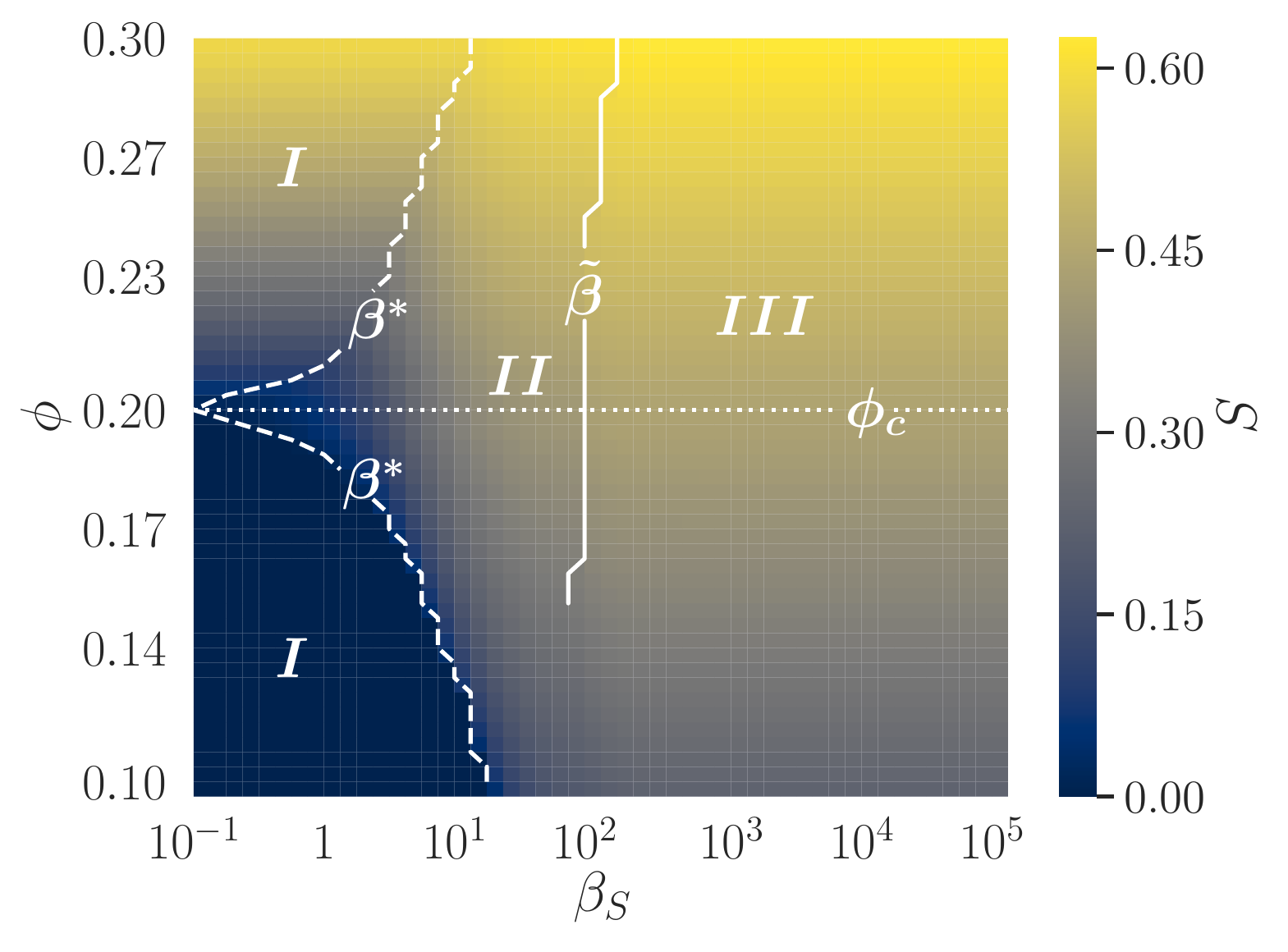}
    \put(0,0){(a)}
  \end{overpic}\\
  \begin{overpic}[width=\columnwidth]{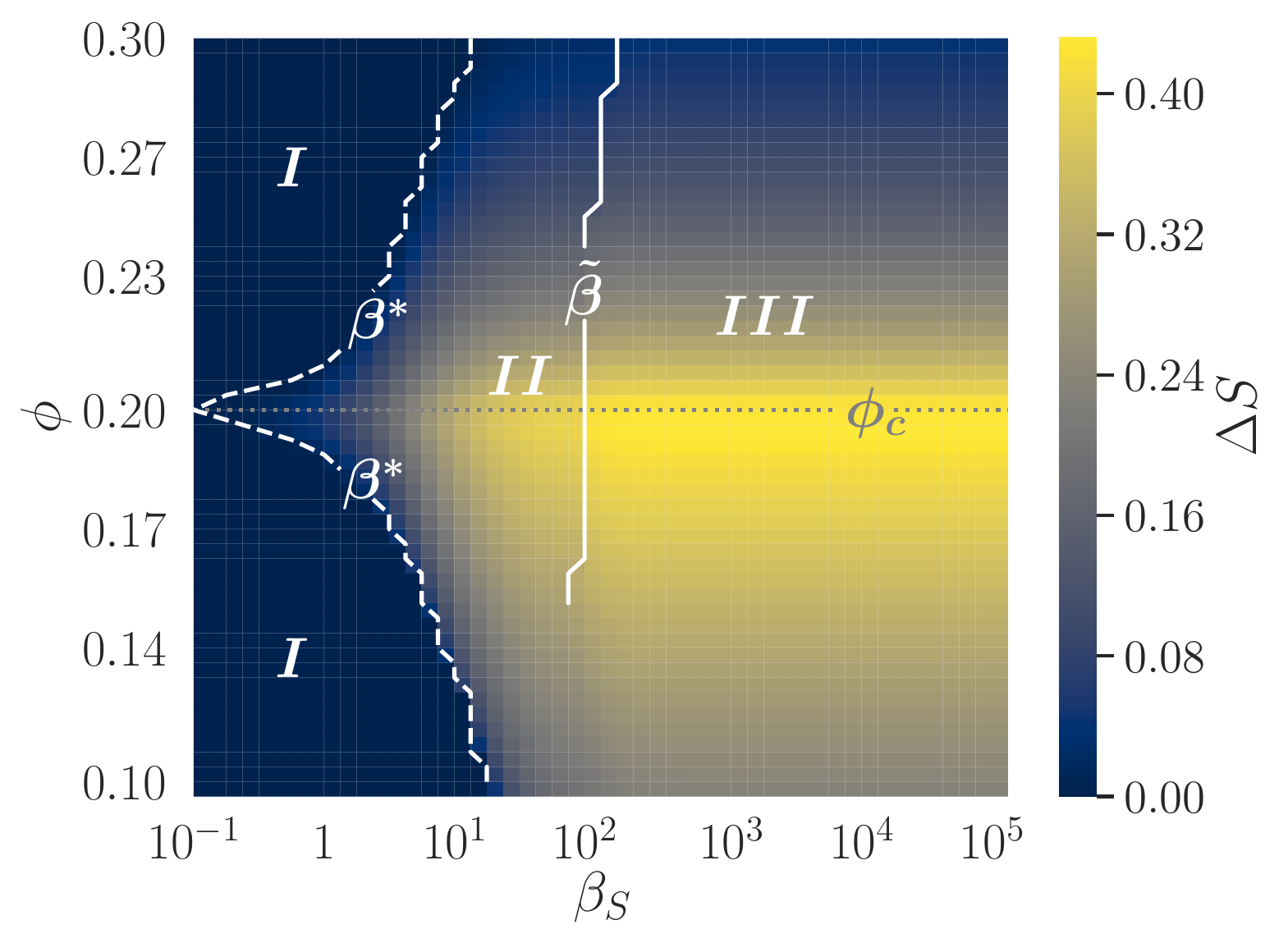}
    \put(0,0){(b)}
  \end{overpic}
\end{tabular}
\caption{(a) Value of the fraction of nodes $S$ which are part of the giant
        connected component as a function of the selective pressure $\beta_S$
        and dilution probability $\phi$. (b) Variation in $S$ with respect to
        the case where no selective pressure is applied as a function of the
        selective pressure $\beta_S$ and dilution probability
        $\phi$.\label{fig:S/S-heatmaps}}
\end{figure}

In Fig.~\ref{fig:S-vs-bS} we show the properties of the obtained models
for $\avg{k} = 5$ and $B=2$ groups. As $\beta_S$ increases, the network
ensemble undergoes two abrupt transitions, where the structure first
changes from fully random ($I$) to a core-periphery structure ($II$),
and finally to an asymmetric bipartite structure ($III$). The
core-periphery structure corresponds to a smaller and denser set of
``core'' nodes which are connected among themselves, and a larger and
sparser set of ``periphery'' nodes which connect mostly to the core
nodes, and not among themselves. The asymmetric bipartite structure is
similar to the core-periphery pattern, but the ``core'' nodes no longer
preferentially connect to themselves, instead they predominantly connect
to the periphery nodes, although they remain a smaller and denser
set. An illustration of these structures can be seen in
Fig.~\ref{fig:S/low-phi-nw-samples}, where we show network samples from
the obtained ensembles. In Figs.~\ref{fig:S/core-sizes} and
\ref{fig:S/periphery-sizes} we also show the size and density of the two
groups as a function of the selective pressure, for different values of
the edge dilution probability $\phi$.

It easy to understand why a core-periphery structure increases the
robustness to random edge removal: the core group corresponds to a
denser subgraph which remains connected with a large probability after
the removal of a given fraction of edges, and the peripheral nodes
benefit directly from this by connecting directly to the core, rather
than among themselves. What is perhaps more surprising is the eventual
onset of the bipartite structure, at which point the core group becomes
so dense that its nodes tend to remain in the giant component even if
they are not connected preferentially among themselves, which would
incur a large entropy cost for no significant additional benefit, but
instead connect mostly to periphery nodes. The latter group tends to
remain connected since its nodes tend to receive multiple connections to
the denser core nodes. (Similar structures to the core-periphery one
encountered here were also seen in similar setups where the robustness
was integrated over all possible dilution values
$\phi$~\cite{peixoto_evolution_2012,priester_limits_2014} as well
different ones based on dynamical robustness against
noise~\cite{peixoto_emergence_2012}, but the onset of the bipartite
structures were not seen in these other cases.)

In most cases, the results tend to change predictably with different values of
the edge dilution probability $\phi$, however a qualitative change in behavior
is seen when we cross the $\phi=\phi_c$ value, where $\phi_c=1/\avg{k}$ is the
critical percolation value for a fully random graph. For $\phi > \phi_c$ a
fully random graph has a nonzero giant component even for $\beta_S=0$, and thus
the progression to core-periphery and bipartite structures proceeds as
discussed above. However, for $\phi < \phi_c$ a fully random graph gets
completely disconnected, and therefore the response of the structural changes
to increasing $\beta_S$ is not continuous, but happens more abruptly, with the
onset of core-group that is typically much denser. We observe also an
interesting behavior for sufficiently large values of $\phi$, where the core
group spans almost the entire network at its onset, with an average degree
coinciding with the whole network. The mechanism driving the network structure
as $\beta_S$ increases appears to be slightly different in this case, as it is
the smaller set of ``periphery'' nodes that end up forming the
smaller group of the eventual bipartite structure.

For $\phi=\phi_c$ we also observe a different behavior, where the onset
of the core-periphery structure ceases to be abrupt, and the change
happens continuously. This seems to indicate that an infinitesimal
optimization of networks that lie on the critical percolation threshold
has an infinitesimal entropic cost (a similar behavior had been observed
previously in the context of Boolean networks optimized against
stochastic fluctuations~\cite{peixoto_emergence_2012}).

A more detailed overview of the combined effect of $\beta_S$ and $\phi$
can be seen in Fig.~\ref{fig:S/S-heatmaps}, which shows both the value
of $S(\beta_S,\phi)$, but also the relative improvement $\Delta
S(\beta_S, \phi) = S(\beta_S, \phi) - S(0, \phi)$ with respect to a
fully random graph. Indeed most of the improvement happens around the
critical value $\phi=\phi_c$.

Changing the value of the imposed averaged degree $\avg{k}$ only shifts
the position of the transitions, which remain qualitatively the
same. The value of the number of groups $B$ does not change at all the
results obtained. Indeed, for any value $B > 2$, we find that it is
possible to merge together two or more groups, without changing the
ensemble properties, until only two groups remain. The structures
identified above are then to be considered the only ones to emerge when
the selective pressure against random removal of edges is the only
driving mechanism.

\subsection{Modularity}

\begin{figure}
  \includegraphics[width=\columnwidth]{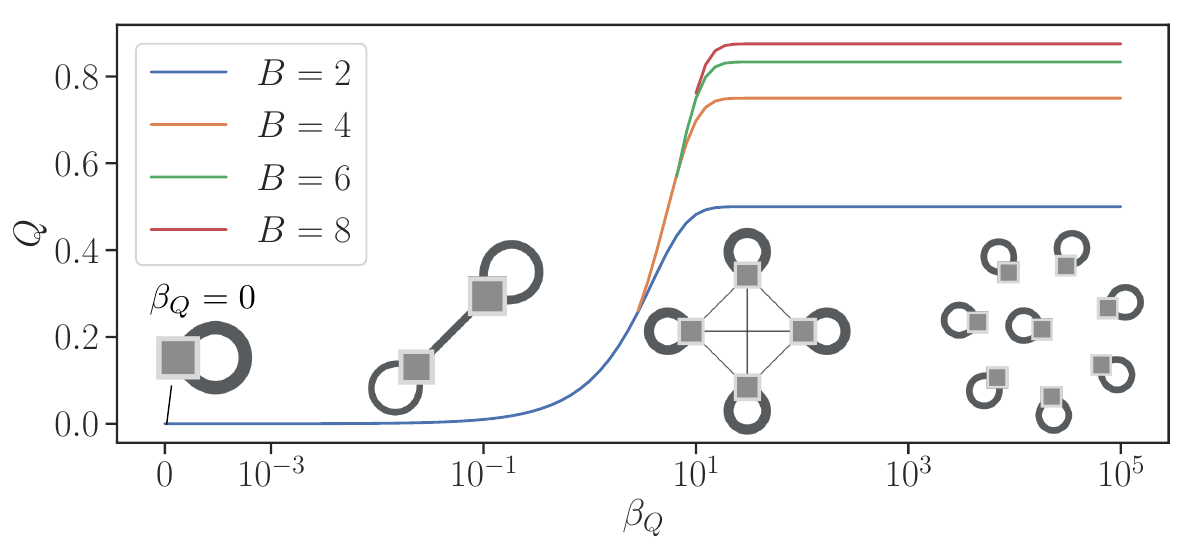}
  \caption{Modularity $Q$ as a function of the selective pressure $\beta_Q$ for
          different choices of the allowed number of groups $B$.}
  \label{fig:Q-vs-bQ-multi-B}
\end{figure}

Some networks tend to be clustered into groups of nodes that are more
connected to themselves than to the rest of the network. This feature
can be beneficial for the adaptability~\cite{kashtan_spontaneous_2005,wagner_road_2007}
and stability~\cite{guimera_origin_2010} of biological systems, and also
to the efficiency of technological systems where these modules are
associated with tasks that can be executed in parallel.

The most typical way to quantify of this kind of assortativity pattern
is via the modularity function~\cite{girvan_community_2002}
\begin{equation}
  Q(\A, \bb) = \frac{1}{2E} \sum_{ij}\left(A_{ij} - \frac{k_ik_j}{2E}\right)\delta_{b_i, b_j},
\end{equation}
where $E$ is the total number of edges in the network and
$k_i=\sum_jA_{ji}$ is the degree of node $i$. The quantity above simply
counts the frequency of edges observed between nodes of the same group,
subtracted by the expected fraction in a fully random graph with the
same degree sequence.  The expected value for modularity for the SBM can
be easily computed as
\begin{equation} \label{eq:Q}
  Q(\m) = \sum_r m_{rr} - m_r^2.
\end{equation}
Note that for completely assortative SBMs with $m_{rs}=\delta_{rs}/B$,
we have $Q(\m)=1-1/B$, so we achieve maximal modularity $Q(\m)\to 1$ for
an infinite number of perfectly isolated groups.

We can include the modularity as a fitness criterion into our framework by
making $R(\w,\m)=Q(\m)$ and coupling with its selective pressure $\beta_Q$, and
proceeding to minimize the free energy
\begin{equation} \label{eq:f-Q}
  \mathcal{F}(\w,\m) = -\beta_Q Q(\m) - \Sigma(\w,\m).
\end{equation}
In Fig.~\ref{fig:Q-vs-bQ-multi-B} we see the value of the modularity
of the network ensemble as a function of the selective pressure
$\beta_Q$. As the selective pressure increases, the network splits
smoothly and progressively into fully symmetric groups of equal size
with a larger number of connections inside each group. For low values of
$\beta_Q$, the results obtained with different number of groups
coincide, and then they start to diverge for higher values. This is
because the actual number of groups populated starts off small and
progressively increases for larger values of $\beta_Q$. Differently from
the percolation scenario considered in the previous setting, we do not
observe abrupt transitions of any kind.

\subsection{Multiple optimization criteria}

We now turn to the situation where we seek to optimize both modularity
and robustness against random edge removal. In principle, this would
amount to a free energy given by
\begin{equation}
  \mathcal{F}(\w,\m) = -\beta_S S(\w,\m) - \beta_Q Q(\m) - \Sigma(\w,\m).
\end{equation}
However, this would mean that the same division of the network used to
compute modularity would also be used to obtain the robustness. But here
we want to be more general, and allow the modularity of the network to
refer to a division that is not necessarily related to the one used to
obtain the robustness. We do so by assuming that the partition used for
the computation of robustness is a subdivision of the one used to obtain
modularity, such that each of its $B_Q$ groups can be further divided
into one, two or more groups, totalling $B_S\ge B_Q$ groups. This
assumption is made without loss of generality, since any two independent
partitions into $B_1$ and $B_2$ groups can always be equivalently
decomposed into one with at most $B_1\times B_2$ groups, which is itself
a subdivision of a smaller one with $\min(B_1,B_2)$ groups. Based on
this, we have the free energy given by
\begin{equation}\label{eq:f-Q+S}
  \mathcal{F}(\w,\m,\bm c) = -\beta_S S(\w,\m) - \beta_Q Q[\m'(\m,\bm c)] - \Sigma(\w,\m).
\end{equation}
where $\bm c= (c_1,\dots,c_{B_S})$ is a hierarchical grouping of the $B_S$
groups, with $c_r\in [1, B_Q]$ being the group membership of
the group $r$ used for the computation of the giant component $S$. The
modularity is therefore computed with the condensed matrix
\begin{equation}
  m_{tu}'(\m,\bm c)=\sum_{rs}m_{rs}\delta_{r,c_t}\delta_{s,c_u}.
\end{equation}
 We stress that for our calculations the identity of the group memberships are
 irrelevant, as we concern ourselves only with the resulting network structures.
 Therefore, we select $B_Q=q$ and $B_S=qk$, where each of the $q$ groups used
 for the computation of $Q$ are subdivided into exactly $k$ groups. Again, this
 comes without a loss of generality, as we do not make any provisions about how
 large each group is, or even if they are occupied at all. Therefore this scheme
 is purely conventional, and does not impose any kind of inherent symmetry or
 network structure on its own. By choosing $q$ and $k$ sufficiently large, we can
 obtain any kind of modular structure used to compute either $S$ or $Q$,
 independently. For our calculations we have used mostly $q=k=2$, which have
 proved sufficient to capture most of the structures seen, but we have
 investigated higher values as well, as we discuss later.

\begin{figure}
  \includegraphics[trim={0 0.10cm 0 0}, clip, width=\columnwidth]{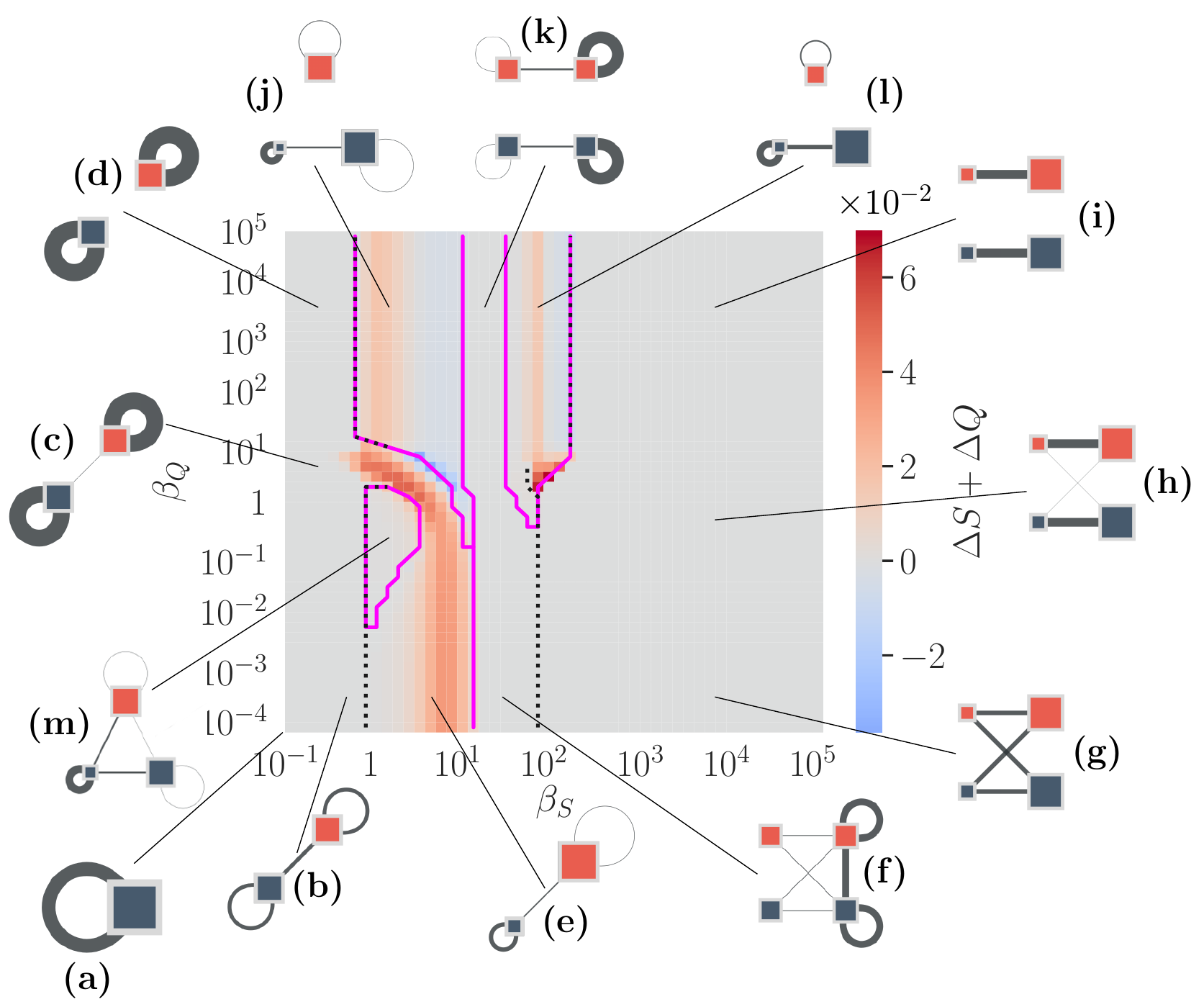}

  \caption{Relative change $\Delta S + \Delta Q$ of the fitness values
    as a function of the selective pressures $\beta_S$ and
    $\beta_Q$. The black dashed lines correspond to transitions
    linked to abrupt changes in the network parameters, the solid
    magenta lines correspond to transitions in which the number of
    groups required to describe the system changes. Schematics of
    the optimized network structures for each region are shown in
    the margins, with each group corresponding to one of the $B_S$
    groups of our model and the color of each group indicating its
    $B_Q$ membership.}
  \label{fig:DQ+DS-heatmap}
\end{figure}

We minimized Eq.~\ref{eq:f-Q+S} for an ensemble of networks with
$\avg{k} = 5$, and edge dilution probability $\phi =
0.21$. Fig.~\ref{fig:DQ+DS-heatmap} shows the relative
changes of the optimization criteria as a function of the selective
pressures $\beta_S$ and $\beta_Q$, where $\Delta S(\beta_S, \beta_Q)$
and $\Delta Q(\beta_S, \beta_Q)$ are defined as
\begin{align}
        \Delta S(\beta_S, \beta_Q) &= S(\beta_S, \beta_Q) - S(\beta_S, 0) \\
        \Delta Q(\beta_S, \beta_Q) &= Q(\beta_S, \beta_Q) - Q(0, \beta_Q)
\end{align}
and represent the relative variations in $S$ and $Q$ induced by the
interplay between the selective pressures with respect to the case in
which we optimized for each constraint in isolation.  As the selective
pressures is changed, we observe a variety of structural phases,
representing diverse combinations of the modular, core-periphery and
bipartite structures encountered previously. The transitions between the
various structures can be either smooth or abrupt. In the latter case,
we can distinguish three types of transitions. The first type of
transition is linked to abrupt changes of the network structure and can
be identified by sudden jumps in the group parameters. The second kind
of transition occurs when the number of groups required to describe the
system changes, but no significant jumps in the group parameters are
observed. Finally, the third type of transition is a mixed transition,
where a change in the number of groups required to describe the system
is accompanied by an abrupt change of the group parameters. Furthermore,
we also observe the presence of synergistic and antagonistic effects,
whereby selecting for one fitness criteria can help (or hinder)
optimizing for the other. We will discuss these effects in more detail
depending on the region where they occur in the phase diagram, as
follows.

\begin{figure}[h!]
  \begin{tabular}{c}
          \begin{overpic}[trim={0, 0.65cm, 0, 0.4cm}, clip, width=\columnwidth]{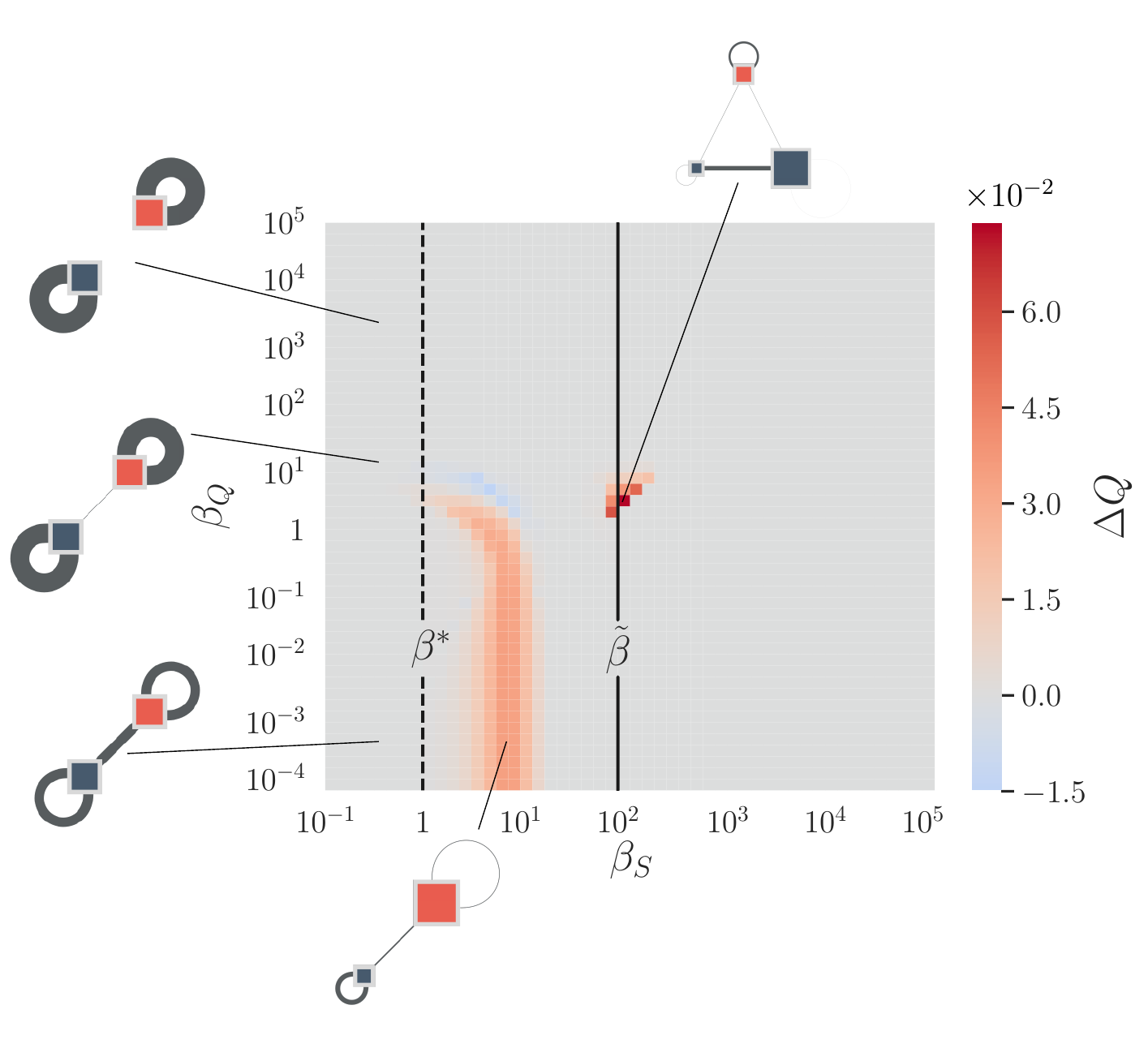}
      \put(0,0){(a)}
    \end{overpic}\\
    \hspace*{0.8cm}
    \begin{overpic}[trim={0, 0.2cm, 0, 0}, clip, width=0.9\columnwidth]{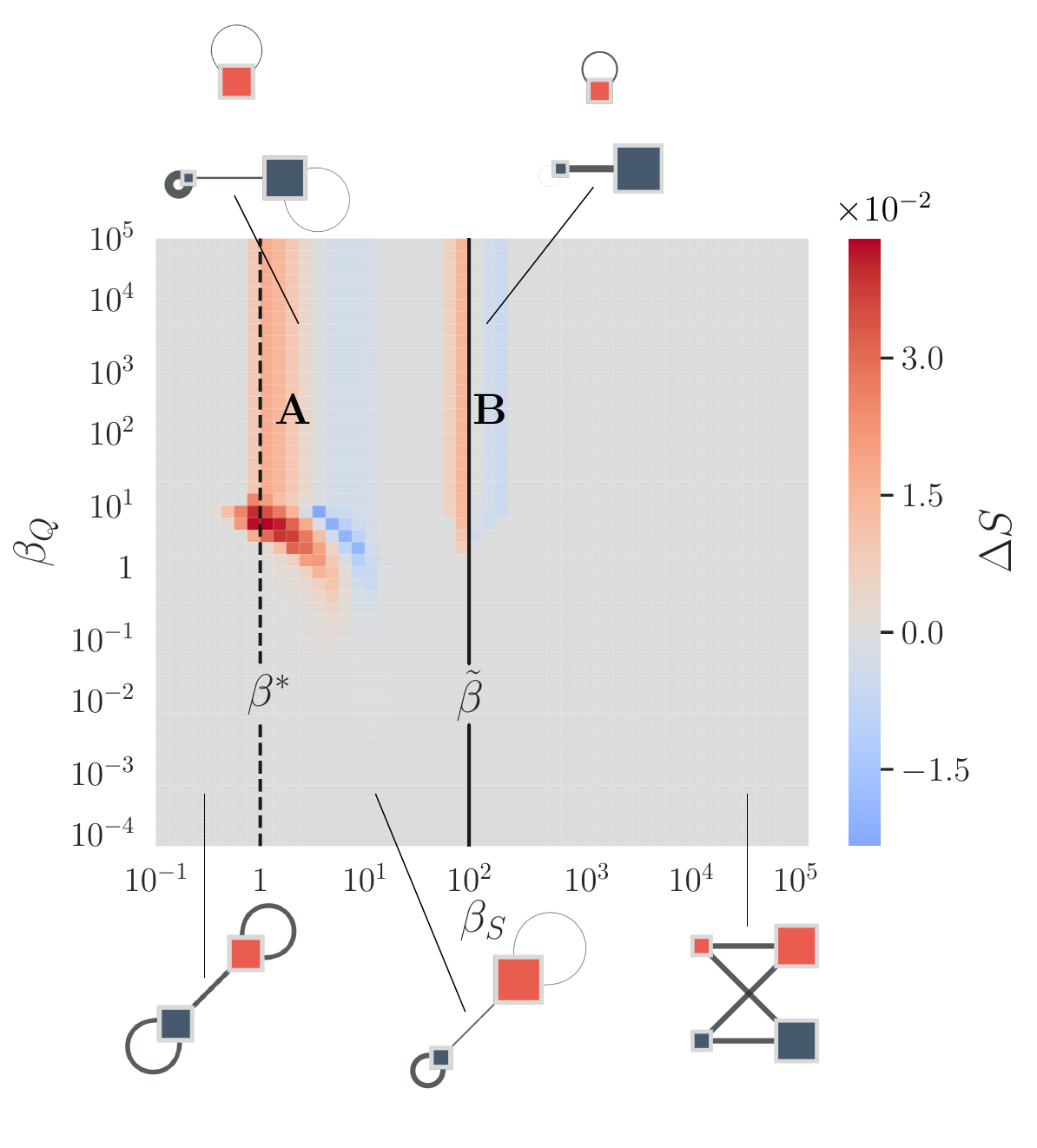}
      \put(0,0){(b)}
    \end{overpic}
  \end{tabular}
  \caption{(a) Change in modularity $Q$ with respect to the case $\beta_S=0$ as
          a function of the selective pressures $\beta_S$ and $\beta_Q$. (b)
          Change in the size of the largest component $S$ with respect to the
          case $\beta_Q=0$ as a function of $\beta_S$ and $\beta_Q$. The dashed
          and solid black lines indicate respectively the values of $\beta_S$
          at which abrupt transitions to core-periphery and bipartite
          structures are observed when optimizing for robustness against random
          edge removal in isolation. Schematics of the optimized structures are
          shown around the margins, where each group corresponds to one of the
          $B_S$ groups in our model and the color of each group indicates its
          $B_Q$ membership.}
          \label{fig:delta-phase-diagrams}
\end{figure}

\subsubsection{Regions in the phase diagram}

\paragraph{The low $\beta_S$ and low $\beta_Q$ regimes:}
For low values of $\beta_S$, we can recover the behavior observed when
selecting for modularity in isolation by varying $\beta_Q$, and the
network structure varies from a random graph, (a) (see
Fig.~\ref{fig:DQ+DS-heatmap}), to increasingly separated and modular
structures, (c) and (d).  Conversely, we note that the behavior observed
when selecting for robustness against random failures in isolation is
not recovered for low $\beta_Q$.  By increasing $\beta_S$ at some fixed
low $\beta_Q$, the network initially follows the expected behavior and
transitions from a random graph, (a), to a core-periphery structure,
(e). However, for high $\beta_S$ the network structure is now described
by a four-group structure composed of two identical and interconnected
core-periphery or bipartite structures, (f) and (g). This symmetric
effect can be understood in terms of modularity. As $\beta_S$ increases,
the selective pressure against random edge removal pushes the network
towards increasingly stronger bipartite structures. Since those
structures have edges running predominantly between different groups,
they would yield negative modularity values. Therefore, by splitting
both ``core'' and ``periphery'' groups each into two random subgroups
used for the computation of modularity, the network can escape the
negative values with negligible entropic cost. Note that, in principle,
one could recover a modularity of zero \emph{and} keep a two-group
structure by simply keeping one of the two $B_Q$ groups empty.  However,
as we can see from Fig.~\ref{fig:Q-vs-bQ-multi-B}, modularity is a
monotonically increasing function of $\beta_Q$, meaning it will only be
zero exactly at $\beta_Q = 0$.  Maintaining a four-group structure where
both $B_Q$ groups are populated allows the network to attain
infinitesimally positive modularity values for $\beta_Q
>0$.

\paragraph{The high $\beta_S$ and high $\beta_Q$ regimes:}
If we increase $\beta_Q$ at some fixed high value of $\beta_S$, we
once again observe that the optimization of modularity causes the
symmetric structures observed above to become less interconnected until
two separate and identical structures coexist, i.e. (g), (h), and (i).
This symmetric pattern effect can be understood as direct consequence of
both optimization criteria competing with each other: since forming a
single mixed core-periphery/bipartite structure would yield low
modularity, the overall structure is mirrored to preserve high fitness
values according to both criteria.

More interesting effects occur if we consider the impact that increasing
$\beta_S$ has at some fixed high value of $\beta_Q$. In this scenario,
we once again observe symmetric structures, see (k) and (i). However, we
also see the presence of regions where the network structure is described
by an asymmetric three-group pattern, see (j) and (l). In these
regions, we again observe the presence of either a core-periphery or
bipartite structure as a result of the selective pressure towards
robustness against random edge removal. The requirement to have a high
fitness for modularity is instead reflected by the presence of an
accompanying and distinct modular structure. This accompanying modular
structure is always denser than a fully random graph and becomes
increasingly dense as $\beta_S$ is increased, suggesting that the
effects of the selective pressure against random edge removal are not
limited to the core-periphery or bipartite structures.

\paragraph{Intermediate regimes:}
For intermediate values of $\beta_S$ and $\beta_Q$, the network
transitions, both smoothly and abruptly, between the same structures
described above. The only difference being the presence of an ``island''
where a three-group pattern again describes the network structure, see
(m). In this region, the structure is that of a core-periphery pattern
in which we now have two peripheries preferentially connecting to a
dense set of core nodes. This structure remains substantially unchanged
if we vary $\beta_S$. By increasing $\beta_Q$, however, one of the two
peripheries becomes progressively smaller and less connected to the
core, and the overall network structure closely resembles the one
observed in (j).

\subsubsection{Synergistic and antagonistic effects}

To better understand the synergistic and antagonistic effects seen in
Fig.~\ref{fig:DQ+DS-heatmap}, it is convenient to consider the
relative variations over $Q$ and $S$ individually, as shown in
Fig.~\ref{fig:delta-phase-diagrams}. Based on this, we consider each
effect in isolation as follows.

\paragraph{Modularity:} Inspecting the diagram for $\Delta Q$ in Fig.~\ref{fig:delta-phase-diagrams}a,
we can see that for low values of $\beta_Q$ and $\beta_S$ the network
structure is essentially that of a fully random graph. By increasing
$\beta_S$, we eventually encounter a synergistic region just above the
$\beta^*$ transition line that exists when $\beta_Q=0$ (see
Fig.~\ref{fig:S-vs-bS}). This indicates that merely transitioning to a
core-periphery structure is enough to guarantee some degree of
improvement in modularity with respect to a random graph. This
synergistic region extends until moderate values of $\beta_Q$,
corresponding to the region in Fig.~\ref{fig:Q-vs-bQ-multi-B} in which
modularity shows a rapid increase. For high values of $\beta_Q$ the
synergistic effects vanish, as we now find ourselves in the region of
Fig.~\ref{fig:Q-vs-bQ-multi-B} where the modularity reaches its
plateau value, and no structural transition can provide an additional
benefit with respect to the case in which we optimize for modularity in
isolation.

What is perhaps more interesting is the small synergistic region in
$\Delta Q$ around the $\tilde{\beta}$ transition line. In this region of
the phase space, the network structure is described by a bipartite
pattern and a separate modular division. It would appear that the
emergence of a bipartite structure --- driven by the selective pressure
towards robustness against edge removal
--- forces more edges to be distributed within their own groups than would be
the case had we selected for modularity alone, thus providing an increased
fitness.

\paragraph{Robustness against random failures:}
In the $\Delta S$ phase space, we observe two principal regions in which
synergistic (antagonistic) effects are present, labelled $A$ and $B$ in
Fig.~\ref{fig:delta-phase-diagrams}.  In region $A$, the network
structure is described by a core-periphery pattern accompanied by an
isolated cluster which is always denser than a fully random graph. This
structure is initially able to provide greater robustness against random
failures than the corresponding two-group core-periphery structures we
observed in Fig.~\ref{fig:S/core-sizes}.  However, it also has a higher
entropic cost, which is accounted by the selective pressure for
modularity, and we observe a synergistic interplay between the two
selective pressures.  This three-group structure displays no significant
changes as $\beta_S$ increases, and, eventually, the evolution of the
core-periphery structures observed in Fig.~\ref{fig:S/core-sizes} can
provide greater robustness. At this point, the selective pressure for
modularity reverses its role by pinning the less optimal three-group
structure in place, and we observe an antagonistic interplay between the
two selective pressures.  Increasing $\beta_S$ even further, we
eventually reach the point where it is more beneficial for the network
to pay a further cost in entropy and split into two symmetric
structures, in exchange for larger mutual fitness.

A similar picture occurs in region $B$, where the network structure is
characterised by a bipartite pattern and an accompanying cluster which
is always denser than a fully random graph. The onset of region $B$
happens for values of $\beta_S \leq \tilde{\beta}$, and the added
bipartitness initially provides an increased fitness against random edge
removal. However, the role of the selective pressure for modularity once
again reverses as soon as $\beta_S > \tilde{\beta}$ and we cross the
bipartite transition line observed when optimizing for robustness
against random edge removal in isolation.

\subsubsection{Increasing the number of groups}
As mentioned at the beginning of the section, it would in principle be
possible to obtain any kind of modular structure by choosing high enough
values of $q$ and $k$. However, the computation of the free energy grows
quadratically with $B_S$, making it computationally expensive to
increase the number of groups used to model the network. Nevertheless,
we have investigated regions of the phase diagram allowing us to probe
in more detail how the allowed number of groups affects the results. Our
findings appear to indicate that increasing the number of groups can
exacerbate the synergistic and antagonistic effects, but does not alter
the regions in which these are observed. However, increasing the number
of groups can potentially give rise to different entanglements of the
core-periphery, bipartite, and modular structures observed above.

\begin{figure}
  \includegraphics[trim={0 0.3cm 0 0}, clip, width=\columnwidth]{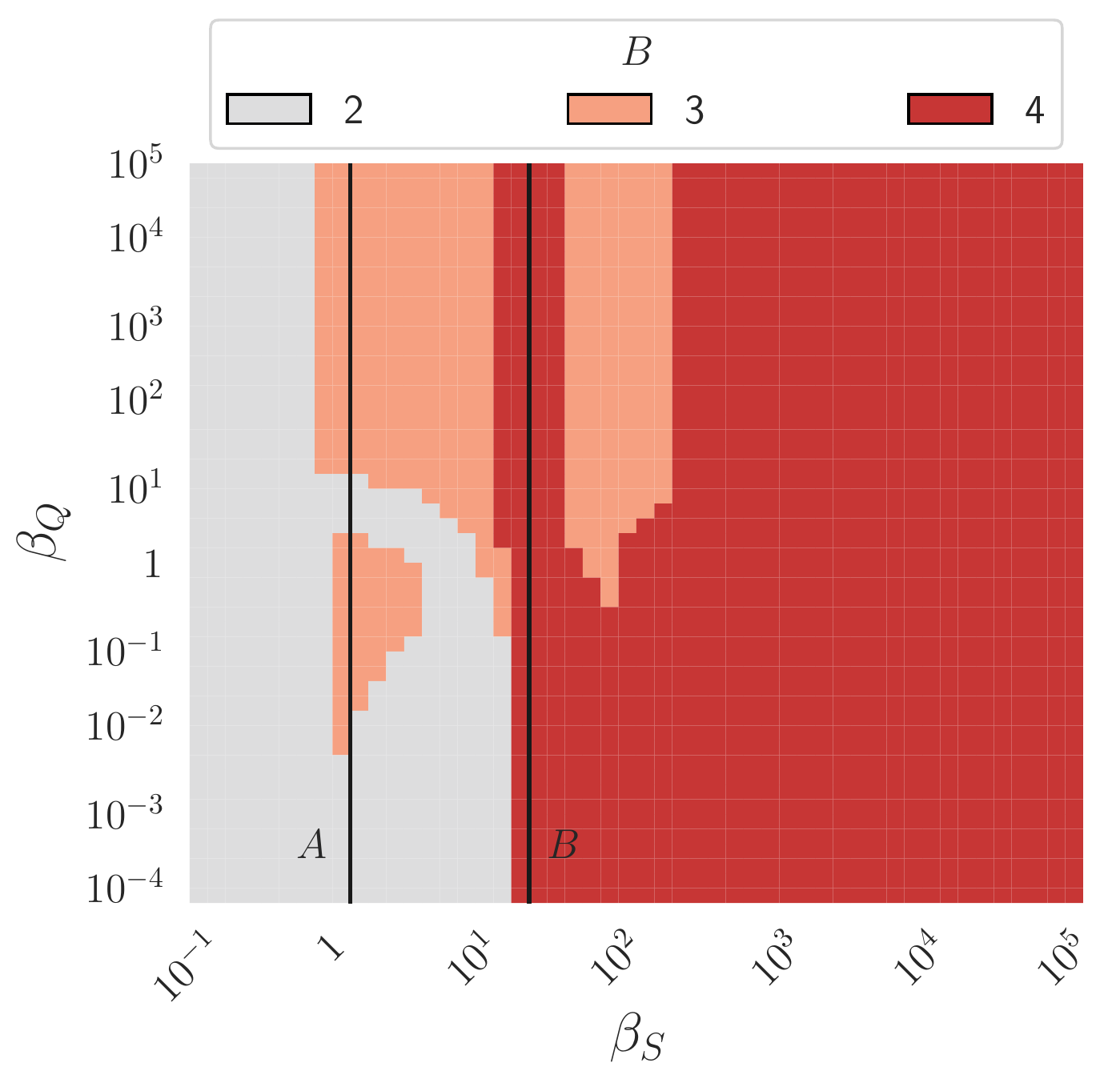}

  \caption{Number of groups required to describe the system as a function
    of the selective pressures $\beta_S$ and $\beta_Q$, up to a maximum
    of $B=4$. The solid black lines indicate the slices $A$ and $B$
    discussed in the text.}
  \label{fig:B-heatmap}
\end{figure}

\begin{figure}
  \begin{tabular}{c}
          \begin{overpic}[trim={0.1cm, 0, 0, 0}, clip, width=\columnwidth]{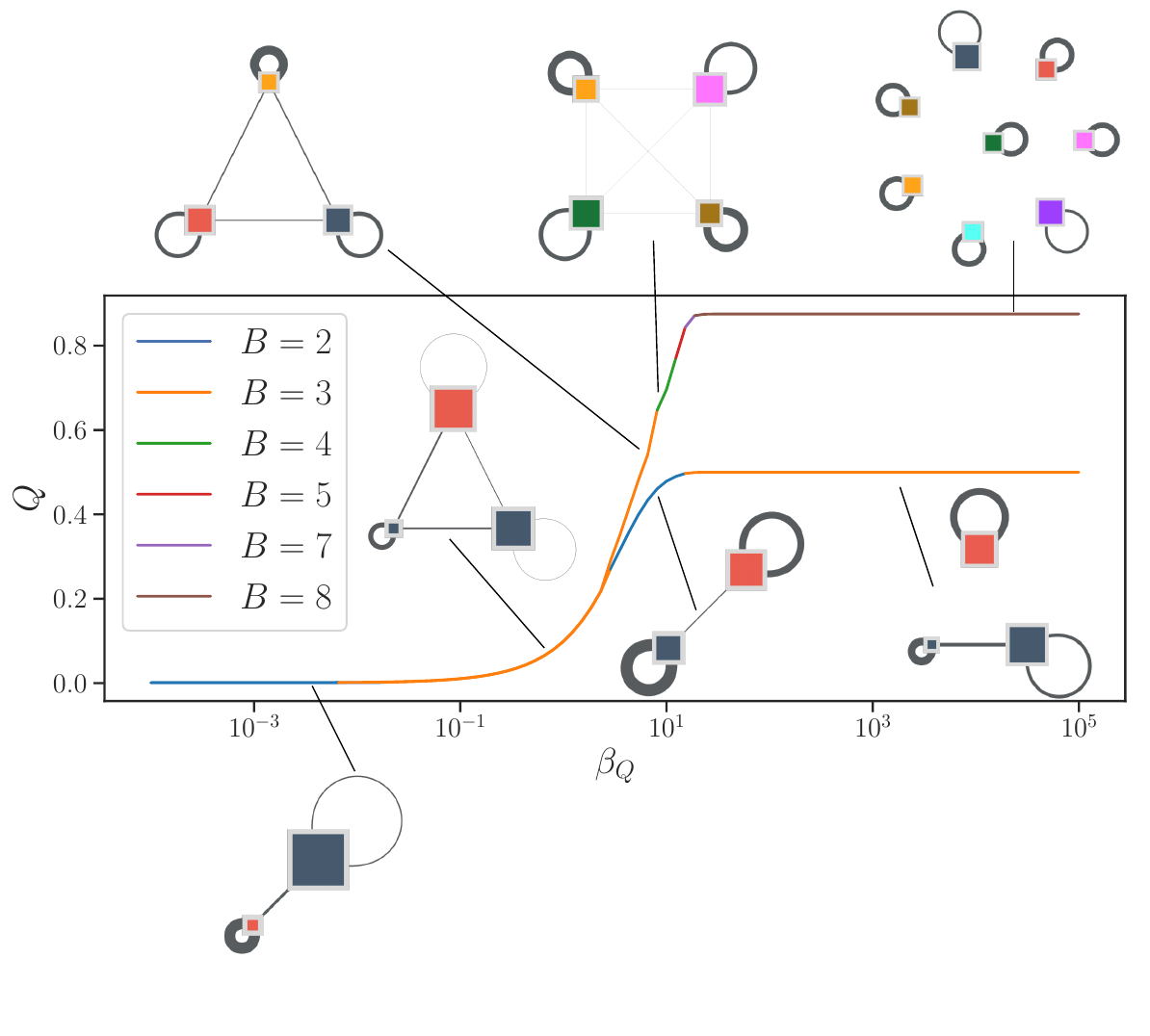}
      \put(0,0){(a)}
    \end{overpic}\\
    \hspace*{0.15cm}
    \begin{overpic}[trim={0.1cm, 0, 0, 0}, clip, width=\columnwidth]{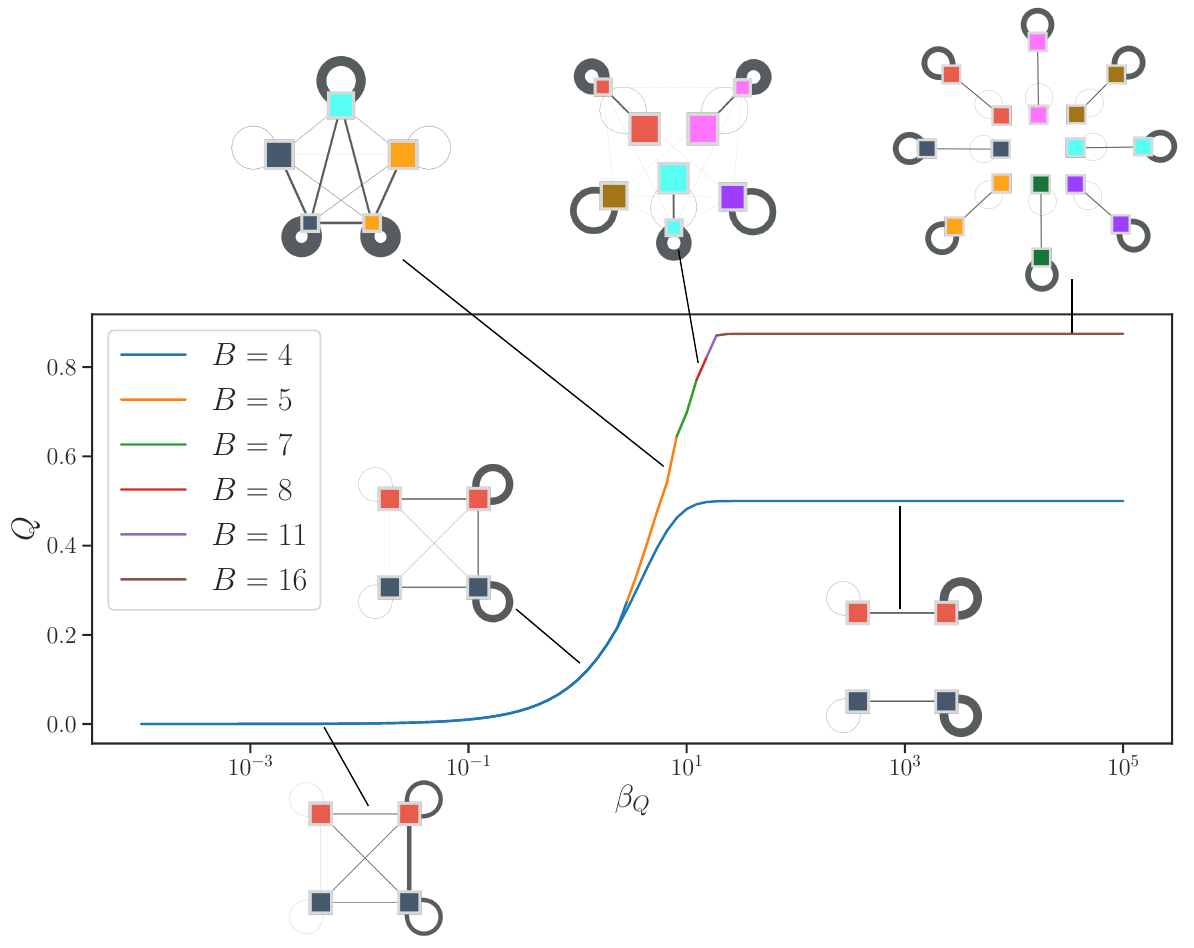}
      \put(0,0){(b)}
    \end{overpic}
  \end{tabular}
  \caption{(a) Modularity as a function of the selective pressure $\beta_Q$ for
          slice $A$. (b) Modularity as a function of the selective pressure
          $\beta_Q$ for slice $B$. The bottom curves display the behavior
          observed in the $q = k = 2$ case, while the top curves
          represent the $q = 8, k = 2$ case. Changes in color indicate a
          change in the number of groups required to describe the system.
          Schematics of the optimized structures are shown in the insets, where
          each group corresponds to one of the $B_S$ groups in our model and the
          color of each group indicates its $B_Q$ membership.}
      \label{fig:B-increase-slices}
\end{figure}

As an example, we consider the two slices at fixed $\beta_S$ shown in
Fig.~\ref{fig:B-heatmap}. For each of these slices, we fix $q = 8$ and
$k = 2$. Fig.~\ref{fig:B-increase-slices} shows a comparison of the
modularity as a function of $\beta_Q$ for both the $q = k = 2$ case
studied above, and this new case with $q = 8$ and $k = 2$. 

For slice $A$, we can see that the two curves coincide for low to moderate
values of $\beta_Q$, with the network structure transitioning from a two-group
core-periphery structure to a three-group core-periphery, where we now have two
peripheries connecting to a dense core group (see (m) in
Fig.~\ref{fig:DQ+DS-heatmap}).  For higher values of $\beta_Q$ the curves
diverge, as the higher value of $q$ in the $q = 8$, $k = 2$ case allows the
network to populate more groups, thus increasing its modularity (and thereby
decreasing the free energy).  The number of groups which are populated increases
with $\beta_Q$, and the network topology is described by interconnected modular
structures which become progressively disconnected from each other as the
selective pressure is raised. We note that, in contrast to what we observed when
we optimized for modularity in isolation, these new modular structures are not
symmetric, with some groups being denser than a random graph and others less so.

For slice $B$, we find ourselves in a region of the parameter space
where the network topology is described by two symmetric core-periphery
structures which get progressively disconnected as $\beta_Q$ is
raised. Once again, the two curves coincide for low to moderate values
of $\beta_Q$, but, as $\beta_Q$ increases, the access to a higher number
of $B_Q$ groups in the $q = 8, k = 2$ case allows for more groups to be
populated, and we observe different entanglements of core-periphery
structures accompanied by isolated clusters, and, for high enough
$\beta_Q$, we once again observe a mirroring effect in which the network
topology is now described by eight symmetric core-periphery structures.

\section{Conclusion}
\label{sec:conclusion}

We have introduced a framework to generate null models of optimized
networks, which allow us to incorporate the effects that selective
pressures toward some predefined set of criteria can have on the
structural properties of the network. A central feature of our approach
is the ability to incorporate an arbitrary number of criteria, allowing
us to analyse more realistic scenarios in which network systems are
subject to multiple interacting selective pressures.

We have applied this framework to analyse the emerging structures in
systems subject to the joint optimization for modularity and robustness
against random removal of edges, which we analysed both in isolation and
in combination. In the case of modularity alone, we showed that by
increasing the selective pressure, we observe network structures which
progressively split into an increasing number of symmetric groups whose
nodes predominantly connect amongst themselves. In the case of
robustness against random failures, we instead identify two phase
transitions in which the network structure transitions first to a
core-periphery pattern, and then into an asymmetric bipartite one. The
core-periphery structure is characterised by a smaller and denser set of
``core'' nodes which connect preferentially amongst themselves and a
larger ``periphery'' whose nodes mostly connect to the core nodes. This
structure allows for higher robustness as the random removal of any edge
is unlikely to disconnect the core, and peripheral nodes remain
connected via the core itself. By increasing the selective pressure
further, the core group eventually becomes so dense that its nodes no
longer require to preferentially connect amongst each other to ensure a
high level of robustness. They instead connect predominantly to the
periphery and we observe an asymmetric bipartite structure.

Finally, by combining both fitness criteria, we observed different
combinations of the above structures, where the core-periphery and
bipartite structures can either appear in duplicate (i.e. we observe two
symmetric core-periphery or bipartite structures) or accompanied by an
additional cluster which ensures high modularity values. Notably, we
observed regions of the parameter space where the interplay between the
selective pressures can have either synergistic or antagonistic effects,
and optimizing for a specific characteristic can either facilitate or
hinder optimizing for the other.

Our results show how the interaction between different selective
pressures can be combined in simple network models, offering a platform
to investigate the effects that different fitness criteria can have on
the emerging network structures. Furthermore, our model parametrization
is the same used to identify modular structure in empirical
networks~\cite{peixoto_bayesian_2019}, and we expect these two
approaches can be eventually combined in order to identify the dominant
driving mechanisms of network formation from network data.

\bibliography{bib}

\end{document}